\documentclass[MSNbibl,nameyear,dvips]{arxstspdf}
\usepackage{flushend}
\usepackage{stfloats}
\usepackage{graphicx}
\volume{26}
\issue{2}
\pubyear{2011}
\firstpage{271}
\lastpage{287}
\doi{10.1214/10-STS349}

\makeatletter
\newtheorem{theorem}{Theorem}

\newcommand{\var}{\operatorname{Var}}
\newcommand{\data}{\mathrm{data}}

\newcommand{\eqref}[1]{(\ref{#1})}

\makeatother

\begin{document}
\begin{frontmatter}

\title{Estimating Random Effects via Adjustment for Density Maximization\thanksref{T1}}
\relateddoi{T1}{Discussed in \doi{10.1214/11-STS349A}
and \doi{10.1214/11-STS349B}; rejoinder at \doi{10.1214/11-STS349REJ}.}

\runtitle{Estimating Random Effects}

\begin{aug}
\author[a]{\fnms{Carl} \snm{Morris}\corref{}\ead[label=e1]{morris@stat.harvard.edu}}
\and
\author[a]{\fnms{Ruoxi} \snm{Tang}}
\runauthor{C. Morris and R. Tang}

\affiliation{University of Harvard}

\address[a]{Carl Morris is Professor, Department of
Statistics, Harvard University, One Oxford Street,
Cambridge, Massachusetts 02138, USA \printead{e1}.
Ruoxi Tang obtained his Ph.D. from Harvard's
Statistics Department and is with a New York investment firm,
Bloomberg L.P.,
731 Lexington Ave.,
New York, New York 10022, USA.}

\end{aug}

%
\begin{abstract}
We develop and evaluate point and interval estimates
for the random effects $\theta_i$, having made observations
$y_i | \theta_i \stackrel{\mathit{ind}}{\sim}
N[\theta_i, V_i], i = 1, \ldots, k$
that follow a two-level Normal hierarchical model.
Fitting this model requires assessing the
Level-2 variance $A \equiv\operatorname{Var}(\theta_i)$ to estimate
shrinkages $B_i \equiv V_i/(V_i+A)$ toward a (possibly estimated)
subspace, with $B_i$ as the target
because the conditional means and variances
of $\theta_i$ depend linearly on~$B_i$, not on $A$.
Adjustment for density maximization, ADM, can do
the fitting for any smooth prior on $A$.
Like the MLE, ADM bases inferences on two derivatives,
but ADM can approximate with any
Pearson family, with Beta distributions being appropriate because
shrinkage factors satisfy $ 0 \le B_i \le1$.

Our emphasis is on frequency properties, which leads to
adopting a~uniform prior on $A \ge0$, which then puts
Stein's harmonic prior (SHP) on the~$k$ random effects.
It is known for the ``equal variances case'' $V_1 = \cdots= V_k$
that formal Bayes procedures for this prior produce
admissible minimax estimates of the random effects, and that the
posterior variances are large enough to provide confidence
intervals that meet their nominal coverages.
Similar results are seen to hold for our approximating
``ADM-SHP'' procedure for equal variances and
also for the unequal variances situations checked here.

For shrinkage coefficient estimation, the
ADM-SHP procedure allows an alternative frequency interpretation.
Writing $L(A)$ as the likelihood of~$B_i$ with $i$ fixed,
ADM-SHP estimates $B_i$ as $\hat{B_i} = V_i/(V_i+ \hat{A})$ with
$\hat{A} \equiv \operatorname{argmax}(A * L(A))$.
This justifies the term ``adjustment for likelihood maximization,''
ALM.
\end{abstract}

%

\end{frontmatter}

\section{Introduction}\label{secintro}
This concerns approximate frequentist, Bayesian,
and objective Bayesian inferences
for a widely applied two-level Normal hierarchical model.
At \mbox{Level-1}, for $i=1, \ldots, k$, unbiased
estimates $y_i$ are observed with means $\theta_i$\vadjust{\goodbreak}
and with known variance $V_i$.
In practice the $\{V_i\}$ usually
are unequal, perhaps with $V_i =\sigma^2/n_i$ and
$\sigma^2$ known or accurately estimated.
Thus
%
%
\begin{equation} y_i | \theta_i \stackrel{\mathit{ind}}{\sim}
N[\theta_i, V_i],\quad i = 1, \ldots, k.
\label{eqlevel1}
\end{equation}
In practice each Level-1 value $y_i$ here
represents a~sufficient statistic or a
summary unbiased estimate based on the $n_i$
observations taken from the $i$th of the $k$ units
(e.g., a hospital, a small area, or a
teaching unit).

Level-2 specifies a Normal model for the random effects $\theta_i$,
each with its own $r$-dimensional predictor variables $x_i$
so that for
$\beta$ and an unknown variance $A\ge0$,
%
%
\begin{equation}
\quad\theta_i | \beta, A \stackrel{\mathit{ind}}{\sim} N[ \mu_i = x_i' \beta
, A],\quad
i=1, \ldots, k.
\label{eqlevel2}
\end{equation}
The case $r=0$ corresponds to $\beta$ fully known and then it
may be convenient to set $\beta=0$ and $\mu_i=0$, WLoG.
If $r\ge1$, $X \equiv(x_1', x_2', \ldots, x_k')'$ as a known $k\times r$
matrix, assumed to have full rank $r$.

The marginal distribution of $y = (y_1, \ldots, y_k)'$, gi\-ven~%
$\beta$ and $A$, and the conditional distribution of $\theta_i$
follow from the above, so that
%
%
\begin{eqnarray}
y_i | \beta, A &\stackrel{\mathit{ind}}{\sim}& N[x'_i \beta, V_i + A],\quad
i=1,\ldots, k,
\label{eqmarginaly}
\\
\label{eqcondtheta}\hspace*{24pt}\theta_i | y_i, \beta, A &\stackrel{\mathit{ind}}{\sim}& N[(1-B_i) y_i + B_i
\mu_i, V_i (1-B_i)],
\nonumber
\\[-8pt]
\\[-8pt]
\nonumber
&&\hspace*{99pt}{} i=1, \ldots, k,
\end{eqnarray}
where $\mu_i \equiv x'_i \beta$, and
$B_i \equiv\frac{V_i}{V_i + A}$ is a ``shrinkage
factor.''\vspace*{2pt}

When $r\ge1$, the vector $\beta$ is assumed throughout to follow
Lebesgue's flat prior on $[0,\infty)$, so
%
%
\begin{equation}
p(\beta, A) \,d\beta \,dA \propto \,d\beta\pi(A) \,dA.
\label{eqlevel3flat}
\end{equation}
Using this flat prior density for $\beta$ is equivalent
to restricted maximum likelihood (REML).
When $\pi(A)$ is proper, the posterior distribution
for this prior is proper (it integrates finitely) if $k \ge r$.
When $\pi(A)$ is improper, a larger $k$ is needed,
with $k \ge r + 3$ sufficing for the
main distributions $\pi(A)$ of interest here.
When $r = 0$, as assumed initially, or when $r \ge1$ and with
$\beta$ integrated out, we can focus on the main issue of
dealing with the (nuisance) variance component $A=\operatorname
{Var}(\theta_i)$
and how to make inferences about the shrinkages $B_i$.

Widely used programs like HLM, ML3 and SAS use MLE/REML methods
to fit this model, while software for fully
Bayesian inferences is available via BUGS and MLwiN
(\citep{rasbbrowgoldyangplewhealwooddraplanglewi01}).
Maximum likelihood and REML obtain an estimate
$\hat{A}$ that maximizes the likelihood function of $A$ (or\vadjust{\eject} mar\-ginal
likelihood in the REML case).
Asymptotically ($k$ large),
maximum likelihood provides optimal estimates of $A$, leading to
convergence of estimates via frequentist and Bayesian approaches.
However, the standard errors assigned by MLE and REML methods to the random
effect estimates and the corresponding interval estimates can
lead to confidence intervals with much smaller than their nominal
confidences, even asymptotically.
This happens with MLE and REML methods not only
because $A$ can be underestimated so that shrinkages are
overestimated, but also because these procedures
do not account for the fact that $A$ has been estimated.

Maximum likelihood and REML estimates of $A$ not
infrequently produce $\hat{A} = 0$, in which case shrinkage MLEs are
$\hat{B}_i = 1$.
Examples occur in every field, as for the 8 schools data
(\citep{gelmcarlsterrubi03}),
and in small area estimation (\citep{Bell99}).
Then, per typical usage, the
variance estimates may be taken to be
$V_i (1-\hat{B}_i) = 0$ when $r = 0$,
leading to zero-width or overly narrow
confidence intervals of $\theta_i$.
As will be seen in Section~\ref{seccoverageprobrisk}, even when $\hat
{A}>0$ and this situation is avoided,
overfitting via MLE and REML can be considerable and
nominal $95\%$ confidence intervals for $\theta_i$
might have true coverages in the 50--80\% range.

The procedures developed here to fit the two-level model above
offer computational ease comparable to maximum likelihood
and REML methods, being based on
differentiating the (adjusted) likelihood function twice.
When $k$ is small or moderate, however, the adjustment provides
much better 
standard errors and interval coverages. ``Better'' coverage is meant
in the Level-2 frequentist sense of averaging over the data and the
Level-2 model (2), for all~fi\-xed~$\beta$, $A$, as illustrated in the
equal variances
case of Figure \ref{figcoveragetheta}, Section~\ref{seccoverageprobrisk}. 

Central to this development is the ADM procedure, ``adjustment for density
maximization'' (\citep{morr88}), albeit not then with the ADM label.
ADM can be used with any Pearson family (Normal, Gamma, Inverted Gamma,
Beta, $F$, $t$ or \mbox{skew-$t$})
to approximate another distribution with a one-dimen\-sional
density. One merely multiplies the density by an adjustment which
is determined by the Pearson family, and then
makes the argmax function
produce the mean, not the mode, of the Pearson distribution.
As seen in \eqref{eqcondtheta},
posterior means and variances of the random effects
are linear functions of the shrinkage factors $B_i$, not of $A$,
so it is desirable to estimate the posterior
mean of $B_i$, and not the mode of $B_i$ or the mean of $A$.
Shrinkage factor distributions are skewed and lie in [0,1],
both of which make a
Beta distribution approximate better than a Normal.
Fitting Beta distributions via ADM is described in Section~\ref{subsecadmapproxbeta}.

Estimating shrinkage factors via ADM
will be seen to reduce to maximizing the posterior density of~$A$
(or the marginalized density, if necessary), after
having multiplied this density by $A$.
This adjustment has several benefits, which
include prevention of estimating $A$ as 0, and overestimating $A$
by just enough to account for the convex dependence of $B_i$ on $A$.
ADM methods have been used successfully before to improve inferences
of random effects in other multilevel models,
as in \citet{chrimorr97} for a Poisson multilevel model.

The main procedure here approximates a formal posterior distribution
stemming from the flat prior $\pi(A) = 1$ on $A \ge0$ in (5).
This flat prior on $A$, in conjunction with \eqref{eqlevel2},
induces Stein's harmonic prior (SHP)
\eqref{eqPriorGoodApproxProportional} on the random effects
(\citep{stei81}) and a minimax admissible estimator. (Stein's prior
on $\bolds{\theta}$ for $ k \ge3$,
$ d \bolds{\theta} / \Vert \bolds{\theta}
\Vert^{(k-2)}$,
is harmonic except at the origin, so it actually is ``superharmonic.''
The shorter term ``harmonic'' is used here for simplicity of discourse.)
The ADM approximations are seen in Section 3 to approximate closely
the exact posterior means and variances of the random effects.
Buttressed with the examples of Sections~\ref{secapproxaccuracy} and
\ref{seccoverageprobrisk}, our assessments show,
by frequency standards, so for all fixed hyperparameters $A \ge0$
and $\beta$, that the ADM-SHP combination outperforms
commonly used MLE and REML procedures for estimating the random effects
\eqref{eqlevel1}--\eqref{eqlevel3flat}.

The ADM approximations of Section~\ref{subsecadmeqvar}
apply to any smooth prior density $\pi(A)$,
including the scale-invariant prior densities $\pi(A)$ on $A$
%
%
\begin{equation}\label{eqsuperharmonicA}
\pi(A) \,dA \propto A^{c-1} \,dA,\quad c>0.
\end{equation}
These receive some specific attention, but
our frequency evaluations are limited
to the special choice in~\eqref{eqsuperharmonicA}
of $c=1$ for which $A \sim \operatorname{Unif} (0, \infty)$.
Stein's harmonic prior not only produces safe frequency procedures for
squared-error point estimation, but the posterior variances
of $\theta_i$ are large enough to serve as a basis for
confidence intervals centered at the posterior means
(\citep{stei81}; Morris \citeyear{morr83,Morris88Purdue}; \citep{chrimorr97}).
Hierarchically, the uniform formal prior $\pi(A)=1$
is suggested by the fact that the renowned James--Stein estimator
is the posterior mean, exactly, if this flat prior is
extended (inappropriately)
to $A \sim\operatorname{Unif} [-V, \infty)$ (\citeauthor{Morris77}, \citeyear{Morris77,morr83}).

Section~\ref{secadm} starts with the ``equal variances
case,'' Stein's setting (\citep{jamestei61}) for which $V_1 =
\cdots= V_k (\equiv V)$.
Although equal variances are unusual in practice,
this situation provides a rich and meaningful structure
that has been studied wide\-ly because of its relative simplicity for
mathematical investigation.
Among other advantages, when $r > 0$ and the
unknown means $\mu_i$ must be estimated,
the equal variances situation allows easy recovery of risks and coverage
probabilities merely by translating these quantities from the simpler
($k - r$)-dimen\-sional situation when shrinkages are toward known
means $\mu_i = 0$.
Also with equal variances, ADM approximations to Bayes rules are easily
developed for the range of scale-invariant priors
\eqref{eqsuperharmonicA}, merely by
solving a quadratic equation for $A$.

Section~\ref{secadm} continues by extending these ADM rules for the
``unequal variance case'' (the variances $V_i$ differ,
as is common in practice).
Section 2.8 introduces a new, more general approximation for the
posterior means and variances, which allows any
$r \ge0$ so that shrinkages can be toward an estimated regression.
With computational and programming methods similar to those of REML,
noticeably more accurate procedures emerge.

Section~\ref{secapproxaccuracy} examines how well ADM methods
approximate the exact Bayes rule.
These approximations are good for small values of $k$
and they become exact as $k \to\infty$.
Even the data analyst who insists on exact computations can
find such approximations useful because of increased speed,
even if only for doing preliminary analyses.

For the case $c=1$ when $A$ is flat, Section~\ref{seccoverageprobrisk}
evaluates the resulting ADM-SHP procedure's
performance in repeated sampling for relative mean squa\-red errors
and for interval coverages.
In the equal variances case, and in the unequal variance examples considered,
nominal coverages are achieved or exceeded for any $k \ge r+3$.
MLE and REML procedures cannot do this.

\section{Adjustment for Density Maximization}\label{secadm} 

This section starts by examining the inadequacy of MLE methods
as a basis for inferences about shrinkage factors $B_i$ and random effects,
and why the ADM approach for shrinkage constants should be better.
For most of this section $r=0$, the dimension of $\beta$, so that
$\beta$ and all $\mu_i \equiv E(\theta_i)$ are assumed known.
Thus, the only unknown Level-2 (nuisance)
parameter is $A$, the between groups variance that governs
the shrinkage factors $B_{i}\equiv\frac{V_{i}}{ V_{i} + A} $.
With $r=0$, (3) and (4) simplify slightly to
%
%
\begin{eqnarray}\label{eqythetaAknown}
y_{i} | A & \sim& N ( \mu_{i}, V_{i} + A ),\nonumber\\
&&\hspace*{-12pt}\mbox{with
shrinkage factor } B_{i} = \frac{V_{i}}{V_{i} + A},\mbox{ and}\\
\hspace*{23pt}\theta_{i} |y_{i},A & \sim& N \bigl( ( 1- B_{i}
) y_{i} + B_{i} \mu_{i}, V_{i} ( 1- B_{i} )
\bigr).\nonumber
\end{eqnarray}
Let $S_{i} \equiv( y_{i} - \mu_{i} )^2 \sim
( V_{i} + A ) \chi^2_{1}$ independently.
$\mathbf{S} \sim(S_{1}, \ldots, S_{k} )'$
is a (minimal, if all $V_i$ differ) sufficient statistic
for $A \geq0$.
Then $\hat{A}_{i}\equiv S_{i} - V_{i}$ for $i = 1, \ldots,k$
are independent unbiased estimates of $A$ with\break
$\operatorname{Var} ( \hat{A}_{i} ) =
2 ( V_{i} + A )^2$. One could average these~$\hat{A}_{i}$,
weighted by the reciprocal of these variances to estimate $A$,
iteratively until convergence, with a negative estimate of $A$ reset to 0.
This produces $\hat{A}_{\mathrm{MLE}}$, the MLE of $A$
(\citep{efromorr75}).

\begin{figure*}

\includegraphics{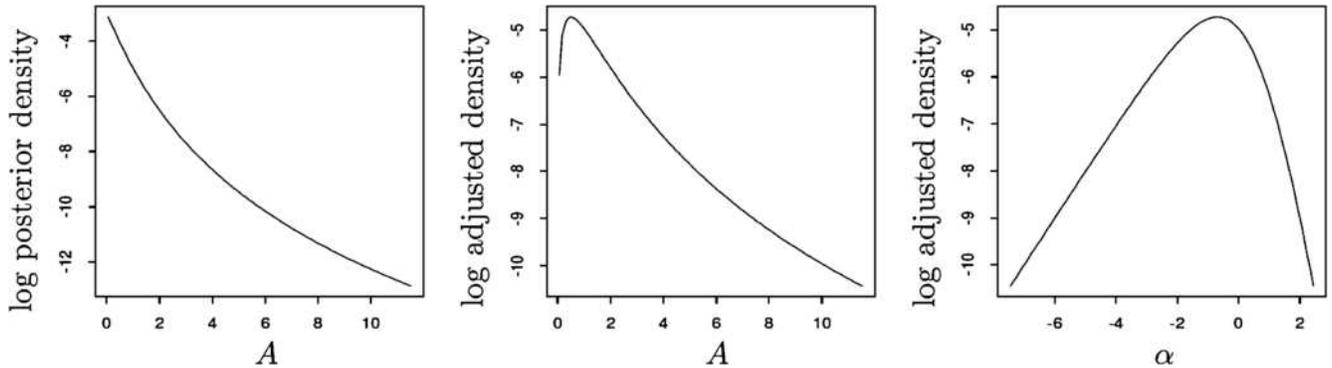}

\caption{An equal variances
example ($V_i=1$) with the MLE on the boundary, $S_{+}=8, k=10, r=0$.
The left panel plots the log posterior density for $A$, which is the
log-likelihood for a flat prior on $A$. The middle panel
plots the log adjusted density against $A$, log$(A*L_0(A))$
in this case, $L_0$-likelihood function (see text and
Section 2.4). The right panel
shows the log adjusted density versus $\alpha\equiv\log A$,
which looks more quadratic.}
\label{figexmleboundary}
\end{figure*}

In the equal variances case,
$S_{+} \equiv\mathop{\sum}_{i=1}^{k} S_{i}$
is complete and sufficient for
$A$, $S_{+} \sim( V + A ) \chi^2_{k}$.
Then
$\hat{A}_{\mathrm{unb}} \equiv
\frac{1}{k} \sum\hat{A}_{i} = \frac{S_{+}}{k} - V$
is unbiased for $A$. Of course, $\hat{A}_{\mathrm{unb}}$ can be
negative, and
$P ( \hat{A}_{\mathrm{unb}} < 0 ) =
P ( \chi^2_{k} \leq k B )$,
where the equal shrinkages are
$B \equiv\frac{V}{V+A}$.
Because $k$ exceeds the median of $\chi^2_{k}$,
$P( \chi^2_{k} \leq k B )>1/2$ if $B$ is near 1 so that
$A$ is near to zero.
This inequality holds for any $k$ if $A \leq\frac{2V}{3k}$,
in which case
$\hat{A}_{\mathrm{unb}} < 0$ and $\hat{A}_{\mathrm{MLE}}=0$
more often than not. This issue of $\hat{A}_{\mathrm{MLE}}$ being zero
or 
quite small has received theoretical attention
at least since \citet{morr83}, and has been recognized
for some time in practice (\citep{Bell99}), 
because its occurrence is not rare. Still, the problem
has yet to be sufficiently recognized so as to be
avoided in practice, and avoided in widely used software.

When $r=0$ the likelihood function is proportional to
%
%
\begin{eqnarray}\label{eqLikelihoodAVdifferent}
L_{0} ( A ) &\equiv&
\Biggl\{\prod_{i=1}^{k} ( V_{i} + A )^{-1/2} \Biggr\}
\nonumber
\\[-8pt]
\\[-8pt]
\nonumber
&&{}\cdot
\exp\Biggl\{ - \frac{1}{2} \sum_{i=1}^{k}
S_{i} / ( V_{i} + A ) \Biggr\}
.
\end{eqnarray}
This is positive at $A=0$ and decreasing near 0
if the $S_{i}$'s are small enough to make
the exponential term be nearly constant.
Then 0 is a local maximum and
if $\hat{A}_{\mathrm{MLE}}=0$
Fisher's information cannot be used to assess
the variance of the MLE.
Furthermore, when $\hat{A}_{\mathrm{MLE}}$ = 0, the MLE of
$\operatorname{Var} ( \theta_{i} | y, A ) =
V_{i} ( 1 - B_{i} )$
also is zero.
An unwary data analyst who uses this for the width of a confidence
interval would assert that $\theta_i = \mu_i$
with arbitrarily high confidence.

The left panel of Figure~\ref{figexmleboundary}
illustrates a case when the logarithm of the
posterior density of $A$, equivalently the
log-likelihood $\operatorname{log}(L_0(A))$
since $A$ has a~flat prior, cannot use Fisher's observed information to
estimate the variance of $A$ since
$\hat{A}_{\mathrm{MLE}} = 0$,
there is no stationary point, and the second derivative
is not negative.
The situation for these data is much improved by using ADM to arrive at
the adjusted log-likelihood in the middle and right panels of
Figure~\ref{figexmleboundary}.

\subsection{Comparing ADM and MLE Methods}\label{secadmestimation}

MLE methods, viewed from a Bayesian (posterior probability)
perspective, amount to finding the posterior
mode of a parameter's distribution and its variance
(reciprocal of observed information) when the parameter has a flat prior
distribution.
Normal distributions are used to approximate the MLE's distribution
based on two derivatives of the log-likeli\-hood.
That works well when the likelihood
is approximately Normal, for example, with large samples,
but it works poorly when likelihoods are quite non-Normal,
as can happen when estimating shrinkage factors.

\citet{morr88}, on approximating posterior distributions,
showed how to fit any prespecified Pearson family (Normal, Gamma, F,
Beta, t, etc.) to a density (but also a likelihood function)
by calculating two derivatives of the ``adjusted'' (posterior)
density function. The adjustment, multiplying by the quadratic
or linear function that
generates the particular Pearson family, makes the maximizer
approximate the mean of
the parameter, and not its mode.
For a nearly symmetric bell-shaped distribution or
likelihood, the Normal is the best Pearson approximation,
the adjustment is a constant.
Then the mode agrees with the mean and the MLE is the ADM.
For skewed likelihoods, the statistician may be
able to choose a better approximating Pearson family,
for example, the Beta family for shrinkage factors.

The following factors compare the ADM and its~fit\-ting process,
perhaps starting with a flat prior on~$A$, with that of the MLE.

\begin{enumerate}[1.]
\item[1.]
Simplicity. An ADM fit is accomplished via a~com\-plexity level
comparable to the MLE, that is, both require two derivatives.

\item[2.] Normality. If a Normal distribution is chosen for the matching
Pearson family, the ADM approach agrees exactly with the MLE,
and the variances in both cases are estimated by
using Fisher's observed information.

\item[3.] Asymptotics. No matter which Pearson distribution is chosen,
ADM provides the same asymptotic inferences (for large $k$) as
the MLE.
This holds because each Pearson family has an asymptotic Normal limit.

\item[4.] Linear expectations. 
While various transformations of a parameter can be considered
for the MLE, ADM targets the mean.
For example, shrin\-kage factors $B_i$ enter linearly
in \eqref{eqcondtheta}, so we approximate their means and variances,
not $A$ or some other function of $B_i$.

\item[5.] Likelihoods?
The ADM procedure could be ter\-med ALM (Adjustment for Likelihood
Maximization), to parallel with MLE language.
ALM and MLE both work best when a version of the parameter
is chosen to represent vague prior information,
giving a relatively flat prior.
We will see that ADM-SHP amounts to maximizing not the likelihood of~$A$,
as the MLE does, but the likelihood after adjustment
via multiplication by $A$.
\citet{lilahiri2010adjusted} proposed using
``adjusted maximum likelihood estimator''
that is identical to ADM if $r=0$.
They showed its advantages in small area estimation
for estimating shrinkages
and for constructing parametric bootstrap prediction intervals.

\item[6.] Multivariate ADM? Adjustments for density
maximization agree with the MLE for approximations via the
Multivariate Normal.
The paucity of non-Normal multivariate Pearson families
restricts ADM's extensions of the MLE to univariate parameters.
However, hybrid extensions are possible, and here we use a
multivariate Normal to approximate the $r$-dimensional
vector $\beta$ and a~Beta distribution for a shrinkage factor.
\end{enumerate}

Given a prior distribution on $A \geq0$, say
$\pi(A ) \,dA$
(proper or not), and still with $r=0$, knowledge of
%
%
\begin{equation}
\hat{B}_{i} \equiv E_{\mbox{\tiny{$\pi$}}}
[ B_{i} | y ]\quad
\mbox{and}\quad
v_{i} \equiv\operatorname{Var}_{\mbox{\tiny{$\pi$}}}
( B_{i} | y )
\label{eqpostmomentsB}
\end{equation}
enables computation of two moments of $\theta_i$,
which with $r = 0$ ($\mu_i$ known) are
%
%
\begin{eqnarray}
\ E [\theta_{i} | y ]
&=& ( 1 - \hat{B}_{i} ) y_{i} + \hat{B}_{i} \mu_{i},
\label{eqpostmeantheta}
\\
\operatorname{Var} ( \theta| y ) &=&
V_{i} (1 - \hat{B}_{i} ) +
v_{i} ( y_{i} - \mu_{i} )^2.
\label{eqpostvartheta}
\end{eqnarray}

The second variance component in \eqref{eqpostvartheta}
often is not represented in MLE applications,
understating variances and encouraging overconfidence.

\subsection{How Maximum Likelihood Can Distort Shrinkage and Random
Effects Inferences}\label{secwhenmlcausesbias}
Each of the following issues can cause overassessment of
the information in the data. This perfect storm can have
serious consequences when $k$ is small or moderate.

\begin{enumerate}[1.]
\item[1.]
Nonlinearity. The posterior means and variances of the
random effects are linear in $B_{i}$, not in $A$.
$B_{i} (A ) = V_{i} / ( V_{i} + A )$
is a convex function of $A$, even if $\hat{A}$ were
unbiased for $A$, one sees, Jensen's inequality, which states that
$B_{i}(E [A | \mathbf{y} ] ) > E [ B_{i} | \mathbf{y} ]$,
indicates that the plug-in shrinkage estimate would be biased too large.
This is why the James--Stein estimator that shrinks
according to $\hat{B}_{\mathrm{JS}}
\equiv\frac{ (k-2)V}{S_{+}}$,\vadjust{\eject} uses the $k-2$ in its numerator, and not
$k$ (as in the MLE), and leads to
smaller mean squared errors than when using MLE shrinkages
$\hat{B} = \frac{k V}{S_{+}}$.

\item[2.]
Boundary limits. Normal approximations to $B_{i}$ put positive
probability outside the boundaries of the interval $[0,1]$.

\item[3.]
Boundary pileup. While
$0 \leq B_{i} \leq1$ is guaranteed, even in the
equal variances case
$V/ ( V + \hat{A}_{\mathrm{unb}} ) =
\frac{kV}{S_{+}} > 1$
is possible.
The MLE cannot exceed 1, but
$\hat{B}_{\mathrm{MLE}} = \operatorname{min} (1,
k V / S_{+} ) = 1$
with positive probability. This pileup happens
despite there being
no prior distribution on $A$, other than
$A = 0$ with certainty, that can allow
$E [ B | \mathbf{y} ] = 1$ for any observation $\mathbf{y}$.

\item[4.]
Skewness. $L_{0} (A )$ tends to be right-skewed,
substantially when the modal value of $A$ is small.
Alternatively, choose a fixed $i$
and replace $A$ by $B_{i}$ in the likelihood by substituting
$A = \frac{1-B_{i}}{B_{i}}V_{i}$ in
$L_{0} ( A )$.\vspace*{2pt} The resulting
likelihood function of $B_{i}$ will be left-skewed.
Approximating such a skewed likelihood by a symmetric (Normal)
distribution overstates the magnitude of $B_{i}$.
A Beta density better approximates an
asymmetric likelihood.

\item[5.] Zero variances. The MLE approach assesses\break
$\operatorname{Var} ( \theta_{i} |  \mathbf{y},A )$
as being $V_{i} ( 1 - \hat{B}_{i,\mathrm{MLE}} )$.
When\break $\hat{A}_{\mathrm{MLE}} = 0$,
this approach in effect attributes
perfect certainty to $A=0$ and that $\theta_i = \mu_i$.

\item[6.] Variance components. Estimating the variance\break of~%
$\theta_i$ by plugging into $V_{i} ( 1 - {B}_{i}) $
overlooks the variance component
$v_{i} = \operatorname{Var} ( B_{i} | y )$ which would
account for the uncertainty in $A$ when estimating $ {B}_{i} $.
Ignoring the term $ v_i (y_i - \mu_i)^2$
amounts to setting $ v_i = 0$.
\end{enumerate}
All six of these biases produces overconfidence.
The unknown variance $A$ is underestimated,
shrinkage~$B_{i}$ is overestimated,
and $\operatorname{Var} ( B_{i} | \mathbf{y} )$ is underestimated.

\subsection{ADM, Adapted to Beta Distributions}\label{subsecadmapproxbeta}
The applications here require approximating
the means and variances of the
shrinkage factors $B_{i}$, $0 \leq B_{i} \leq1$.
Beta distributions are constrained to \mbox{[0, 1]}, so
are the obvious approximating Pearson distribution.
Consider an exact Beta distribution for $B$
with $B \sim\operatorname{Beta} ( a_{1}, a_{0} )$ and density
%
%
\begin{equation}
\qquad f( B ) \,dB =
\frac{ \Gamma(a_{1} ) \Gamma( a_{0} )}
{ \Gamma( a_{1} + a_{0} )}
B^{a_{1}-1} ( 1 - B)^{a_0-1}\,dB.
\label{eqBetaDensity}
\end{equation}
Maximizing\vspace*{2pt} over $B$ gives
$\hat{B} = \frac{a_{1}-1}{ a_{1}+a_{0}-2}$,
the mode (if~$a_{1}, a_{0} \geq1$), not the mean.
The ``adjustment'' for the Beta distribution maximizes the product
$( B ( 1- B ))f(B)$, giving $\hat{B} = \frac
{a_{1}}{a_{1}+a_{0}}$,\vspace*{2pt} the mean of the
$\operatorname{Beta} (a_{1},\break  a_{0} )$ distribution.
Maximizing a Beta density after multiplying by
$B ( 1- B )$ produces the mean, not the mode.

Now let
%
%
\begin{eqnarray}\ell(B) & = &
\log\{ B(1-B)f(B ) \} \\
& = & a_{0} \log B + a_{1}\log(1-B). 
\label{eqAdjustmentLogBeta}
\end{eqnarray}
This is a concave function, maximized uniquely at a~point interior
to (0,1). We have \mbox{$\ell'(B) \!=\!
\frac{a_{1}}{B} \!-\!
\frac{a_{0}}{1-B} \!=\! 0$} at
$\hat{B} = \frac{a_{1}}{a_{1}+a_{0}}$.
Then
%
%
\begin{equation}\label{eqSecondDerivAdjustmentLogBeta}
\qquad - \ell''(B )|_{B = \hat{B}}
= \frac{a_{1}}{\hat{B}^2} +
\frac{a_{0}}{ ( 1 - \hat{B})^2}
= \frac{a_{1}+a_{0}}
{ \hat{B}( 1- \hat{B} )}.
\end{equation}
Thus, given $\hat{B}$ and
$-\ell''(\hat{B} ) > 0$
allows one to recover~$a_{1}$ and $a_{0}$ via
$a_{1}+a_{0} = -\ell'' ( \hat{B} ) \cdot\hat{B}( 1- \hat{B})$
and $a_{1} = \hat{B}( a_{1}+a_{0})$.

If $f(B )$ is a Beta$(a_{1}, a_{0})$ density,
exactly, then
%
%
\begin{eqnarray} E (B ) &=&
\hat{B},\nonumber\\
\label{eqMomentsOfAdjustmentBeta}
v &\equiv& \operatorname{Var} (B ) =
\frac{\hat{B}( 1- \hat{B})}
{a_{1}+a_{0}+1}\\
&=& \frac{\hat{B}( 1- \hat{B})}
{ 1 + \hat{B}( 1- \hat{B} )
(-\ell''( \hat{B} ))}.\nonumber
\end{eqnarray}
If a density $f(B)$ is not exactly
Beta but it lies near to a Beta density, the ADM
approach proceeds similarly, based on two derivatives of
log$( B(1-B)f(B))$, and
approximates
$E [ B ] =
\int_{0}^{1} B f(B )\,dB$ by $\hat{B}$, the
maximizer of this adjusted density. The variance $\operatorname{Var}
(B)$ is approximated by \eqref{eqMomentsOfAdjustmentBeta},
starting with
%
%
\begin{equation}
\ell(B ) \equiv
\log\{ B ( 1- B ) f(B) \}
.
\label{eqlogadjustedbetaadm}
\end{equation}
That is, ADM for a Beta approximation first finds $\hat{B} =
\operatorname{argmax} ( \ell(B ))$.
Then it determines $-\ell''( \hat{B} )$
and uses that to approximate
$\operatorname{Var} (B )$ by
$\frac{\hat{B} ( 1 - \hat{B} )}
{1 + \hat{B} ( 1 - \hat{B} )
(-\ell''( \hat{B} ))}
.$
This Beta distribution approximation to a density on [0,1]
is exact if the original density is a Beta exactly, and it will
be a good approximation if the match is close. Its asymptotic
accuracy can be evaluated favorably (\citep{morr88}, with discussion).

It is useful when fitting shrinkages $B_i = B_i(A)$ to
re-express the results just outlined in terms of $A$, or
equivalently in terms of its logarithm $\alpha=$ log$(A)$,
being sure to include the Jacobian in the posterior density.
Instead of using derivatives of $-\ell(B )$,
the ``invariant information'' will be
calculated, defined\vadjust{\eject} by
%
%
\begin{equation}
\operatorname{inv.info}\equiv- \frac{ d^2 \ell(B)}{ d \{ \operatorname{logit}
(B ) \}^2}\bigg|_{B=\hat{B}}.
\label{eqInvariantInfo}
\end{equation}
The derivative $d \operatorname{logit} (B ) / dB =
d \log( \frac{B}{1-B}) / dB = 1 /\break(B(1-B))$,
which gives
%
%
\begin{eqnarray}\label{eq2ndderivlBlogit}
&&\frac{d^2 \ell(B )}{d \{ \operatorname{logit} (B)\}^2}
\nonumber
\\
&&\quad=B^2 ( 1- B)^2
\ell''(B) \\
&&\qquad{}+
B( 1- B )
( 1 - 2B ) \ell'(B ).\nonumber
\end{eqnarray}
As $\ell'( \hat{B} ) = 0$, we have
$\operatorname{inv.info}=
( \hat{B} ( 1- \hat{B} ))^2\cdot\break
(-\ell''( \hat{B} ) )$.

$\!\!\!$Thus, if $f(B)$ is (nearly) a Beta density
$B \!\sim\!\operatorname{Beta}( a_{1}\!,\break a_{0} )$,\vspace*{2pt} then
$E[ B ] =
\frac{a_{1}}{a_{1}+a_{0}} = \hat{B}$ with
$\hat{B} =  \operatorname{argmax} (\ell(B))$,
and the (approximate) variance is
%
\begin{eqnarray}\label{eqBetaInvariantVar}
\operatorname{Var}(B) &=&
\frac{\hat{B} ( 1- \hat{B} )}
{ a_{1}+a_{0}+1}
\nonumber
\\[-8pt]
\\[-8pt]
\nonumber
&=&
\frac{ (\hat{B} ( 1- \hat{B} ))^2}
{ \operatorname{inv.info} + \hat{B}( 1- \hat{B})} .
\end{eqnarray}

Use of this invariant information is especially valuable
because of the identity
%
%
\begin{eqnarray}\label{eqInvarIdentity}
-\frac{ d^2\ell(B)}
{d\{ \operatorname{logit} (B) \}^2}
&=& -\frac{ d^2 \ell(B (A))}
{d \{ \log(A) \}^2}
\nonumber
\\[-8pt]
\\[-8pt]
\nonumber
&=&
- \frac{ d^2 \ell( B (A(\alpha)))}{
d \alpha^2 }.
\end{eqnarray}
This follows from $d\{\operatorname{logit}(B)\} \!=\!
d \log( \frac{V}{A}) \!=\! - d\alpha$
with $\alpha\equiv\operatorname{log}(A)$.
The invariant information
is the negative second derivative with respect to $\alpha$
of $\ell_{2}(\alpha)$, being the log density
written as a function of $\alpha$:
%
%
\begin{eqnarray}\label{eqInvarianceOfInvariantInfo}
\operatorname{inv.info}& =&
-\frac{d^2\ell(B)}
{ d \{ \operatorname{logit}(B)\}
^2}\bigg|_{B=\hat{B}}
\nonumber\\
&=& - \frac{d^2\ell(B(A))}
{ d ( \log(A))^2}
\bigg|_{A= \hat{A}} \\
&=&
-\frac{d^2 \ell_{2}(\alpha)}{ d \alpha^2 }\bigg|_{\alpha=
\hat{\alpha}}.\nonumber
\end{eqnarray}
Thus, inv.info agrees with Fisher's observed information, but only if
the parameter is $\alpha\equiv\operatorname{log}(A)$.

\subsection{ADM for Estimating Shrinkage Constants}\label{subsecadmshrinkage}
Now return to the Normal model with $r=0$ and likelihood function
$L_{0}(A)$.
Suppose $A \geq0 $ has a prior density
$\pi(A)$, not necessarily proper,
and consider the shrinkage coefficient for
component $i$, $1 \leq i \leq k$,
$B_{i} = \frac{V_{i}}{V_{i}+A}$. The posterior
density for $B_{i}$, given $\mathbf{y}$, is proportional
to $L_{0}(A) \pi(A)\,dA \equiv
f(B_{i}) \,dB_{i}$, where
$A = V_{i}(1-B_{i}) / B_{i}$ and
$dA = -V_{i} \,dB_{i} / B_{i}^2$.
Then\break
$f(B_{i}) \equiv L_{0}(A)
\pi(A)V_{i} / B_{i}^2$ is proportional to
the density of $B_{i}$. To apply ADM, define
%
%
\begin{eqnarray} \ell_{0} (B_{i}) & \equiv&
\log\bigl(B_{i}(1-B_{i}) f( B_{i}) \bigr) \\
& = & \log( A \pi(A)L_{0}(A)
) \equiv\ell(A).
\label{eqEll0Bi}
\end{eqnarray}
Still thinking of $A$ as a function of $B_{i}$,
%
%
\begin{equation}
\frac{d \ell(A )}{d B_{i}} = \frac{dA}{dB_{i}} \frac{ d
\ell(A )}{ dA} =
\frac{-V_{i}}{B_{i}^2}\ell'(A).
\label{eqSec24PartialsOfAB}
\end{equation}

The following theorem summarizes what has just been demonstrated about
the ADM approximation by a Beta distribution for $B_i=V_i/(V_i+A)$,
starting with a posterior density on $A$ that is proportional to
$L_{0}(A)\pi(A)$.
\begin{theorem}
\label{adm-theorem1approx}
Given a prior density $\pi(A)$ and a~likelihood function
$L_{0}(A)$, the ADM procedure for a~Beta distribution
approximates the first two posterior
moments of $B_{i}$ as
%
%
\begin{equation}
E [ B_{i} | \mathbf{y} ] =
\hat{B}_{i} = \frac{V_{i}}{V_{i}+\hat{A}},
\label{eqEBgivenyisVoverVplushatA}
\end{equation}
where $\hat{A} = \operatorname{argmax}( \ell(A ) )$,
$\ell(A) \equiv
\log(A \pi(A) L_{0}(A) )$,
and

%
\begin{equation}
v_{i} \equiv\operatorname{Var}( B_{i} | \mathbf{y} ) =
\frac{(\hat{B}_{i} ( 1 - \hat{B}_{i}))^2}
{ \operatorname{inv.info} + \hat{B}_{i}( 1- \hat{B}_{i})}
,
\label{eqVBgivenyEquation}
\end{equation}
with $\operatorname{inv.info}\equiv- \ell''(\hat{A}) \hat{A}^2$.
\end{theorem}

Neither $\hat{A}$ nor the invariant information
depends on $i$ or on $V_i$.

\subsection{Priors for Good Frequency Performance}\label{subsecpriorgoodfreqperformance}
Admissible rules, which are Bayes and extended Bayes rules
(per the ``fundamental theorem of decision theory''), can
provide good frequency properties if they
are based on priors that let the data speak.
One way to do that restricts to scale invariant improper
priors $\pi(A) \,dA = A^{c-1} \,dA$, $0 < c \leq1$.
As discussed earlier,
given $k$, these priors with $c \geq c_{k} > 0$ ($c_k<1/2$, but not too
small) produce estimators of~$\theta_{i}$ whose\vadjust{\goodbreak}
posterior means are minimax estimators
for squared-error loss in the equal variance setting, so that for all
vectors $\bolds{\theta}$ (fixed),
%
%
\begin{equation}
E \sum_{i=1}^{k}\bigl\{
\bigl( 1- \hat{B} (S_{+}) \bigr)y_{i} - \theta_{i} \bigr\}
^2/V < k,
\end{equation}
%
\begin{equation}
\hat{B}(S_{+})\equiv E \biggl[
\frac{V}{V+A} \Big| S_{+} \biggr].
\label{eqSteinSense}
\end{equation}
The choice $c=0$, so $\pi(A)\,dA=dA/A$, puts
essentially all mass at $A$ nearly 0,
making $\hat{B}(S_{+}) = 1$
with certainty, no matter what the data say.
This choice must be avoided, but sometimes it is not.
As $c$ increases,
shrinkages $\hat{B}(S_{+})$ decrease. For
$c=1$ and for some smaller values, down to $c_k$,
minimax and admissible estimators result.

Our preference $A \sim\operatorname{Uniform}(0, \infty)$ is
equivalent to Stein's harmonic prior, that is,
for $\bolds{\theta} \in\mathbb{R}^{k}$, $k \geq3$, the (improper)
measure on $\bolds{\theta}$ is seen to be
$ d \bolds{\theta} / \Vert \bolds{\theta}
\Vert^{(k-2)}$.
This is the density of $\bolds{\theta}$ if, independently for
$i = 1, \ldots, k$,
$\theta_i | A \sim N (0, A)$ and
$A \sim\operatorname{Unif} [0, \infty)$, as seen from
%
%
\begin{equation}
\int_{0}^{\infty}
e^{-{1}/{2} \Vert\bolds{ \theta}\Vert^2/A}\frac{dA}{A^{k/2}}
\propto\Vert\bolds{ \theta}\Vert^{2-k}.
\label{eqPriorGoodApproxProportional}
\end{equation}
This prior with $c=1$, that is, $A \sim\operatorname{Unif}[0, \infty
)$,
is strongly suggested in the equal variance case by~the fact that the
James--Stein shrinkage constant $\hat{B} = \frac{k-2}{S_{+}}$ is precisely
the posterior mean $E [ \frac{V}{V+A}| S_{+}]$
if~$A~\!\sim\break\operatorname{Unif} [ -V, \infty)$.
Lopping off the impossible part where $A < 0$ leads to $A \sim
\operatorname{Unif}
[0, \infty)$ (\citep{Morris83Box}).
That the James--Stein estimator is asymptotically optimal for large
$\Vert \theta\Vert$ further suggests its
use, that is, choosing $c=1$.
Still in the equal variances case,
some values of $c < 1$, for example $c=1/2$, shrink harder,
which lowers the summed mean squared error if
$\Vert \bolds{\theta}\Vert^2$
is suspected not to be large.
Experience with this flat prior on $A$ has borne out its good frequency
properties in a variety of situations, also including for unequal variances.
Supporting evidence is given in Sections 3 and 4.

\begin{figure*}

\includegraphics{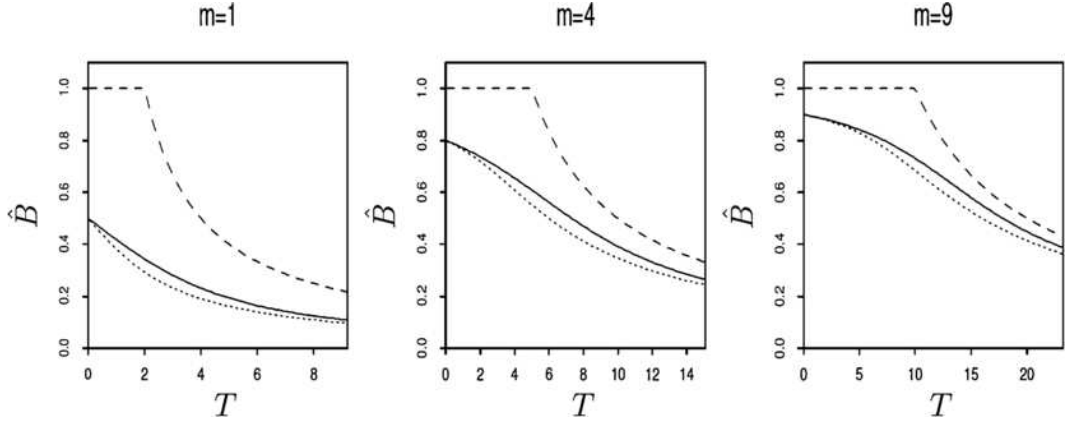}

\caption[Plot of $\hat{B}$ versus $T \equiv RSS/(2V)$ from three different
methods]{Plot of
$\hat{B}$ versus $T \equiv S_{+}/2V$ from three different
methods, with $m=1, 4, 9$ ($k=4,10,20$), respectively.
The solid line is from the exact calculation, the dotted
line is from ADM, and the dashed line is the MLE.}
\label{figBvsT}
\end{figure*}

\subsection{Exact Moments for the Uniform Prior in the Equal Variances
Case}\label{subsecexactmomentseqvar}

The exact posterior means and variances of
$B\!=\!V/\break(V+A)$ for $c=1$, $A$ being uniform (\citep{Morris83Box}),
are as follows.
Denote $m \equiv(k-r-2)/2$, so $m = ( k-2)/2$ when $r=0$.
If $r > 0$, the dimension of $\beta$, then the one can shrink
toward the $r$-dimensional fitted
subspace determined by $\hat{\beta} \equiv(X'X)^{-1} X' y$.
In the ($k - r$)-dimensional space orthogonal
to the range of $X$, shrinkage is toward the $0$-vector.
We\vadjust{\eject} therefore can focus on that $k-r$ subspace with $r=0$ and~$k$
replacing $k-r$ (or think of shrinkage\vadjust{\goodbreak} as toward a known,
fixed vector $\mu$ as here). Now with $y_i \sim
N(\mu_{i}, V+A)$, let $S_{+} \equiv\sum_{i=1}^k (y_i-\mu_{i})^2$, and
let $T \equiv S_{+}/2V$. The James--Stein estimate is
$\hat{B}_{\mathrm{JS}} \equiv m/T = ( k-r-2 )V/S_{+}$.
Let $M_m(T)$ be the moment generating
function of a $\operatorname{Beta} (1, m)$ distribution at $T$, a confluent
hypergeometric function (\citep{abrasteg64}),
%
%
\begin{eqnarray}
\qquad\quad M_m(T) &\equiv& \int_0^1 \exp[(1-B)T]\,dB^m
\nonumber
\\[-8pt]
\\[-8pt]
\nonumber
&=&\Gamma(m+1)T^{-m}\exp(T)P(\chi^2_{2m}\le2T).
\end{eqnarray}
Then (\citep{Morris83Box}),
%
%
\begin{eqnarray}
\label{eqBhatexact}
\hat{B}_{\mathrm{exact}} & \equiv &E[B|S]\nonumber\\
&=& \frac
{m}{T}\bigl(1-1/M_m(T)\bigr)
\nonumber
\\[-8pt]
\\[-8pt]
\nonumber
&= &\frac{(k-r-2) V}{ S_{+}}\\
&&{}\cdot \frac{ P(\chi
^2_{2m+2} \leq S_{+} / V )}{ P ( \chi^2_{2m} \leq S_{+}/V
)}, \nonumber\\
\label{eqvarBexact}
v_{\mathrm{exact}}&\equiv &\var(B|S)
\nonumber\\
&= &\frac{1}{m} \hat
{B}_{\mathrm{exact}}^2 - (\hat{B}_{\mathrm{JS}}
- \hat{B}_{\mathrm{exact}})\\
&&\hspace*{28pt}\qquad\cdot\biggl(1-\frac{m+1}{m} \hat{B}_{\mathrm{exact}}
\biggr).\nonumber
\end{eqnarray}
With $r = 0$, it follows that
%
%
\begin{eqnarray}
\label{eqthetahatexact}\hat{\theta}_{\mathrm{exact},i} & \equiv& E[\theta_i | \mathbf
{y}]
\nonumber
\\[-8pt]
\\[-8pt]
\nonumber
&=& (1-\hat{B}_{\mathrm{exact}}) y_i +
\hat{B}_{\mathrm{exact}} \mu_{i},\\
\label{eqvarthetahatexact}
s_{\mathrm{exact}, i}^2 & \equiv& \var(\theta_i | \mathbf{y})
\nonumber
\\[-8pt]
\\[-8pt]
\nonumber
& =&
V (1 - \hat{B}_{\mathrm{exact}})
+ v_{\mathrm{exact}} (y_i - \mu_{i})^2.
\end{eqnarray}
The elegance of these formulas for the equal variances case
is striking. Unfortunately, this disappears in the unequal variances case
that invariably arises in practice, which motivates the search
for relatively simple alternatives to exact calculations.

\subsection{ADM for Shrinkages, Equal Variances Case}\label{subsecadmeqvar}

Maximum likelihood estimates have optimal
asym\-ptotic properties, but the small and moderate sample sizes ($k$)
that arise in hierarchical modeling applications may be too small
for the MLE to perform well.
The mode of $A$, or more relevantly of $B$, may be quite
inadequate approximations to the posterior mean that
corresponds to a flat prior that makes\vadjust{\eject} the likelihood
agree with the posterior density.
Figure~\ref{figexmleboundary} provides
a simple example for equal variances, scaled for a sample size
$k=10$ with shrinkage toward zero ($r=0$)
and a sufficient statistic $S_{+} = 8$.
$S_{+} = 8$ is the mode of a $\chi^2_{10}$ distribution,
and also is the largest value of $S$ that makes the James--Stein shrinkage
estimate $\hat{B}_{\mathrm{JS}} = 1$.
Likelihood graphs like this are not uncommon in practice,
even when unequal variances occur.
The right-most panels, which have made an adjustment
to the likelihood, make it possible for two derivatives to
capture the distribution, whereas there is no hope of this
with the unadjusted left panel.

Figure~\ref{figBvsT} plots estimated shrinkages $\hat{B}$
against $T = S_+/2V$, for values of $k = 4, 10, 20$, each panel
showing three different estimation methods:
the exact shrinkage estimate for the flat harmonic prior $c=1$, SHP
(solid curve);
the ADM approximation to the same prior (dotted);
and the MLE $ =\operatorname{min}(1, (m+1)/T) = \operatorname{min}
(1,k/S_+)$.
The MLE shrinks much more heavily than the other two methods
when $T$ (or $S_+$) is small.
The ADM shrinkage curves are fairly close to
the exactly computed expected shrinkage in each case,
but are slightly more conservative.

When $\beta$ is unknown so that $r > 0$,
the marginal distribution of $A$
is gotten by integrating $\beta$ out of
the joint posterior density of $\beta$ and $A$
(which is done in the next section, and extended to unequal variances).
The marginal density is neatly written in this equal variances case
in terms of the sum of squared residuals, $S_+ \equiv\sum_i(y_i-
\hat{y}_i)^2$
and $\hat{y} \equiv X\hat\beta$ as
%
%
\begin{eqnarray}\label{eqposteriorA}
p(A|y) &\propto&(V+A)^{-{(k-r)}/{2}}
\nonumber
\\[-10pt]
\\[-10pt]
\nonumber
&&{}\cdot\exp
\biggl\{ - \frac{S_+}{2(V+A)} \biggr\} \pi(A).
\end{eqnarray}
For $\pi(A) \propto A^{c-1}$, the logarithm of the adjusted density
(multiplying by $A$) is
%
%
\begin{eqnarray}\label{eqlposteriorA}
\ell_2 (A | y) &\equiv& c \log A - (m+1) \log(V+A)
\nonumber
\\[-8pt]
\\[-8pt]
\nonumber
&&{}- \frac{TV}{V+A},
\end{eqnarray}
$T \equiv S_+/2V$.
With no covariates, $r=0$, this equation continues to hold
with $m = (k-2)/2$.

Now,
%
%
\begin{eqnarray}\label{eqconvexqdtAhat}
&&\frac{d \ell_2 (\alpha) }{d \alpha} \nonumber\\
&&\quad= A \frac{d \ell_2}{d A}
\nonumber\\
&&\quad =  - \bigl((m+1-c) A^2\\
&&\quad\qquad{} - (2c+T-m-1)V A - c
V^2\bigr)\nonumber\\
&&\qquad{}/(V+A)^2.\nonumber
\end{eqnarray}
The numerator of (\ref{eqconvexqdtAhat}) is a convex quadratic
function of $A$ (with $m+1-c>0$) which is negative at $A=0$.
It therefore has two real roots, one negative and
unacceptable.
The positive root is the ADM estimator $\hat{A}$.
Then,
%
%
\begin{eqnarray}\label{eqsolutionBhat}
\hat{B} &\equiv&\frac{V}{V+\hat{A}}
\nonumber
\\[-8pt]
\\[-8pt]
\nonumber
&=&
\frac{2(m-c+1)}{T+m+1+\sqrt{(T-m-1)^2 + 4cT}}.
\end{eqnarray}
Note that $\hat{B}$ is monotone decreasing in $T$ and that
$\hat{B}$ reaches its maximum, $1 - c/(m+1) < 1$ at $T=0$.
Shrinkage is bounded away from $100\%$ if $c > 0$, for example,
if $c=1$ and $r=0$ the maximum shrinkage is $(k-2)/k$.
These shrinkages\vadjust{\goodbreak}
$\hat{B}$ decrease as $c$ increases and as $c\rightarrow0$ in \eqref{eqsolutionBhat}, $\hat{B}\rightarrow\operatorname{min}((m+1)/T,1)$.
Of course $c = 0$ is not allowed because then the
posterior guarantees $100\%$ shrinkage, no matter what the data say.

Define $\alpha\equiv\operatorname{log}(A)$ and $\hat{\alpha} \equiv\log
\hat{A}$.
Then for any $c$,
the invariant information${} = {}$inv.info satisfies
%
%
\begin{eqnarray}
\operatorname{inv.info} &=&
- \frac{d^2
\ell_2}{d \alpha^2} \bigg|_{\alpha=\hat{\alpha}}
\nonumber
\\[-8pt]
\\[-8pt]
\nonumber
&=& m(1 - \hat{B})^2 + \hat{B}^2 + (1-c)(1-2\hat{B}).
\end{eqnarray}
Matching the first and second derivatives of the two densities
(i.e., of the adjusted density and of a $\operatorname{Beta} (a_1, \break a_0)$ density)
gives
\[
a_1 = \frac{\operatorname{inv.info}}{1-\hat{B}},\quad  a_0 = \frac
{\operatorname{inv.info}}{\hat{B}},
\]
and this Beta distribution has variance
%
%
\begin{eqnarray}\label{eqvBeta}
v & = & \frac{\hat{B} (1-\hat{B})}{a_0 + a_1 +1}
\nonumber
\\[-8pt]
\\[-8pt]
\nonumber
& = & \frac{\hat{B}^2 (1-\hat{B})^2}{m (1 - \hat{B})^2 + (1-c) +
(2c-1) \hat{B}}.
\end{eqnarray}
When $c=1$, the ADM approximations in this equal variances case to the
posterior moments of $B=V/\break(V+A)$ are
%
\begin{eqnarray}
\hat{B}&=&\frac{2(k-r-2)V}{S_{+}+kV+\sqrt{(S_+-kV)^2+8S_+V}},\label{eqBhatapprox}\\
v&=&\frac{\hat{B}^2(1-\hat{B})^2}{m(1-\hat{B})^2+\hat
{B}}.\label{eqvarBhatapprox}
\end{eqnarray}

\begin{figure}

\includegraphics{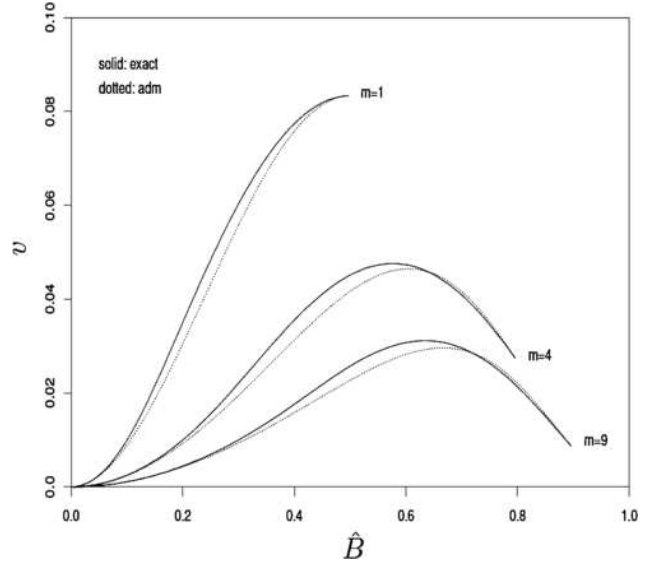}
\vspace*{-3pt}
\caption{Plot of $v$ versus its own $\hat{B}$ from two different
methods. The solid line is from the exact method, that is, formulas~%
\protect\eqref{eqBhatexact} and~\protect\eqref{eqvarBexact}, and the dotted
line is from the approximate method, formulas~\protect\eqref{eqBhatapprox} and
\protect\eqref{eqvarBhatapprox}.}
\label{figBvsvhat}
\vspace*{-5pt}
\end{figure}

For the SHP case $c=1$ in Figure~\ref{figBvsT}, $\hat{B}$ is plotted
as a function of $T$, showing that the ADM estimate
of $B$ shrinks slightly less than the exactly computed $B$, while it matches
exactly at $T = 0$, and asymptotes to the exact value for large $T$. The
MLE produces much larger shrinkages.

Figure~\ref{figBvsvhat}, as in Figure~\ref{figBvsT}, also
shows graphs for the SHP ($c=1$) and with curves
for $m=1,4,9$ (e.g., if $r=0$, then for $k = 4, 10, 20$).
It reveals that the ADM approximation
to $v$ corresponds well with the exact posterior variance of a
shrinkage factor,
each as a function of its own shrinkage $\hat{B}$.
In both cases the shrinkage $\hat{B}$
decreases monotonically as the sufficient
statistic $S_+$ rises.
Figure~\ref{figBvsvhat} shows ADM's excellent
ADM approximation of the exact variance, and that it becomes
exact as $T$ nears 0 (where maximal shrinkage in both cases is
for $\hat{B}$ = $m/(m+1) = (k-r-2)/(k-2)$).\vadjust{\goodbreak}

For any $r \ge0$ in this equal variance case, the preceding
estimates of the shrinkages and of their variances provide the
following estimates of the means and the variances of the
random effects $\theta_i$ in terms of the ADM approximations to the
posterior moments $\hat{B}$ and
$\hat\beta\equiv(X'X)^{-1}X'y$:
%
\begin{eqnarray}
\hat\theta_{i} &\equiv&\hat{E} (\theta_i | y)
\nonumber
\\[-8pt]
\\[-8pt]
\nonumber
&=& (1-\hat{B})
y_i + \hat{B} x_i' \hat\beta,
\label{eqthetahatapprox}
\\
\label{eqvarthetahatapprox}
s_{i}^2 &\equiv&\widehat{\var}(\theta_i | y)
\nonumber\\
&=& V(1-\hat{B}) + V\bigl(x'_i (X'X)^{-1} x_i\bigr)\hat{B} \\
&&{}+ v
(y_i-x'_i\hat{\beta})^2.\nonumber
\end{eqnarray}
Note that $s_{i}^2$ depends on $i$ by increasing proportionally
to the squared residual, as one would expect because mis-estimation
of $B$ hardly matters when $(y_i-x'_i\hat{\beta})^2$ is small.
These results are seen most easily by using a least squares
regression predictor in the $r$-dimensional range space of
$X$, and shrinking to $0$ in the ($k-r$)-dimensional orthogonal
subspace. The extension to the unequal variance case, which is next,
is more complicated.

\subsection{The Unequal Variances Case With Regression} \label{subsecuneqlvarcase}

An ADM approach to fitting our general model 
starts
by integrating out the $\{\theta_i\}$ to get, in matrix notation,
%
%
\begin{equation}
y|\beta, A\sim N_k(X\beta, D_{V+A}),
\label{eqgenmodygivenbetaA}
\end{equation}
where $D_{V+A}\equiv\operatorname{diag}(V_i+A)$ 
is a $k$-by-$k$ diagonal matrix.
With $\beta$ having a flat prior on $R^r$, standard
calculations with \eqref{eqgenmodygivenbetaA} lead to
%
%
\begin{equation}
\qquad\hat\beta_A\equiv
E(\beta|y,A)=(X^{\prime}D_{V+A}^{-1}X)^{-1}X^{\prime}D_{V+A}^{-1}y.
\label{eqgenmodbetahat}
\end{equation}
With $A$ known, $\hat\beta_A$ is at once both the posterior mean and the
weighted least squares estimate of $\beta$.
The full distribution, given $A$, is
%
%
\begin{equation}\label{eqgenmoddistrbeta}
\beta|A,y
\sim N_r(\hat\beta_A,(X^{\prime}D_{V+A}^{-1}X)^{-1}).
\end{equation}

The objective is to make inferences about the vector
$\bold\theta= (\theta_1, \ldots, \theta_k)$
with conditional distribution
%
%
\begin{eqnarray}\label{eqgenmoddistrtheta}
\theta|\beta,A,y
&\sim& N_k\bigl((I-B_A)y+B_AX\beta,
\nonumber
\\[-8pt]
\\[-8pt]
\nonumber
&&\hspace*{58pt}(I-B_A)V\bigr).
\end{eqnarray}
This is \eqref{eqcondtheta} in matrix notation, with $I$ the
\mbox{$k$-by-$k$}~iden\-tity matrix,
$V\!\equiv\!\operatorname{diag} (V_1,\ldots,V_k)$
and $B_A\!\equiv\operatorname{diag} (B_i \!= V_i/(V_i+A)).$
Integrating out $\beta$, with help from \eqref{eqgenmodbetahat},
it follows that
%
%
\begin{eqnarray}\label{eqgenmoddistrthetaintegrateoutbeta}
\hspace*{7pt}\theta|A,y&\sim& N_k\bigl((I-B_A)y+B_AX\hat\beta_A,
\nonumber
\\
&&\hspace*{22pt}(I-B_A)V\\
&&\hspace*{22pt}{}+V^{1/2}B_A^{1/2}P_AB_A^{1/2}V^{1/2}\bigr),\nonumber
\end{eqnarray}
where in \eqref{eqgenmoddistrthetaintegrateoutbeta}
$P_A$ is a $k\times k$ projection matrix of rank~$r$,
%
%
\begin{equation}
P_A\equiv D_{V+A}^{-1/2}
X(X^{\prime}D_{V+A}^{-1}X)^{-1}X^{\prime}D_{V+A}^{-1/2}.
\label{eqgenmodPA}
\end{equation}

When $A$ has prior density element $\pi(A)\,dA$, the posterior density of
$A$, given $y$, follows:
%
%
\begin{eqnarray}\label{eqgenmodpostA}
\quad p(A|y)&\propto&|D_{V+A}|^{-1/2}|X^{\prime}D_{V+A}^{-1}X|^{-1/2}
\nonumber
\\
&&{}\cdot\operatorname{exp}
\bigl(-\tfrac{1}{2}(y-X\hat\beta_A)^{\prime}\\
&&\hspace*{24pt}{}\cdot D_{V+A}^{-1}(y-X\hat\beta_A)\bigr).\nonumber
\end{eqnarray}

The logarithm of this adjusted posterior density, with
$\alpha=\log(A)$, is
%
%
\begin{eqnarray}\label{eqgenmodlpostA}
l(\alpha)&=&\log(A\pi(A)
)\nonumber\\
&&{}-\frac{1}{2}\sum_{1}^k\log(V_i+A)
\nonumber
\\[-8pt]
\\[-8pt]
\nonumber
&&{}-\frac{1}{2}\log
|X^{\prime}D_{V+A}^{-1}X|\\
&&{}-\frac{1}{2}(y-X\hat\beta_A)^{\prime
}D_{V+A}^{-1}(y-X\hat\beta_A).\nonumber
\end{eqnarray}
Denote $\hat\alpha\!\equiv\!\operatorname{argmax}(l(\alpha))$,
set $\hat A\!=\!\operatorname{exp}(\hat\alpha)$,
and defi\-ne $\operatorname{inv.info} \equiv-l{}^{\prime\prime}(\hat\alpha).$
Then the ADM approximation, with
$\hat B_i \equiv V_i/(V_i+\hat A)$, is
$B_i\equiv\frac{V_i}{V_i+A}\sim\operatorname{Beta}$
with approximate mean $E(B_i) = \hat B_i=V_i/(V_i+\hat A)$
and\vadjust{\eject} variance $ v_i = \operatorname{Var}(B_i) = \{\hat
B_i(1-\hat B_i)\}^2/\{\operatorname{inv.info}+\hat B_i(1-\hat B_i)\},$
both moments depending on the prior $\pi(A)$.
Maximizing $\ell(\alpha)$ and determining its second derivative at
$\hat\alpha$, the negative of the invariant information,
can be done by numerical methods, by Newton's method (which requires
matrix derivatives), or by other means that include an EM technique
available in \citet{tang02}.

Given $\hat A$ and the values $\{\hat B_i, v_i\}, i=1,\ldots,k$, one
could insert $\hat A$ into \eqref{eqgenmoddistrthetaintegrateoutbeta}
to estimate both posterior moments of the $\theta_i$.
However, that underestimates the variance
and makes no use of the $\{v_i\}$, so we proceed as follows, leading
to a main theorem.

Define $\hat\beta$ as $\hat\beta_A$ evaluated at $\hat A$ and
$\hat y \equiv X\hat\beta.$
Then from \eqref{eqgenmoddistrthetaintegrateoutbeta},
and approximating
$\hat\beta_A$ by $\hat\beta$,
%
%
\begin{eqnarray}
E(\theta_i|A,y)&\doteq &y_i-B_i(y_i-\hat y_i),
\label{eqgenmodexpthetaiAy}
\\
\operatorname{Var}(E(\theta_i|A,y))&\doteq& v_i(y_i-\hat y_i)^2.
\label{eqgenmodvarexpthetaiAy}
\end{eqnarray}

To minimize complications in making our final approximations
to $E(\theta_i|y)$ and $\operatorname{Var}(\theta_i|y)$,
we neglect variations of $\hat\beta_A$ in \eqref{eqgenmodbetahat}
and $P_A$
in \eqref{eqgenmodPA} as $A$ varies around $\hat A$.
This is exact in the equal variances case
because both $\hat\beta_A$ and $P_A$ do not depend on $A$,
and it will be nearly true if the $\{V_i\}, i=1,\ldots,k$ differ only
slightly.
With unequal variances both $\hat\beta_A$ and \eqref{eqgenmodPA}
involve weights that depend on
$\{V_i+A\}$.
If $\hat A$ is near $A$, as happens when $k$ is
large, then $\frac{V_i+\hat A}{V_i+A}$ is near 1.
With data, one can evaluate
%
%
\begin{equation}
\operatorname{Var} \biggl\{\biggl(\frac{V_i+\hat A}{V_i+A}\biggr)\Big\vert y
\biggr\}=\operatorname{Var}\biggl(\frac{B_i}{\hat B_i}\Big\vert y\biggr)=\frac{v_i}{\hat
B_i^{2}}.
\end{equation}
These variances may be acceptably small, and
$v_i/\hat B_i^{2}$ diminishes as $1/k$ as
$k\longrightarrow\infty.$

\begin{theorem} Assume the model \eqref{eqlevel1}, \eqref{eqlevel2},
and the prior in \eqref{eqlevel3flat}.
Write $\hat B_i$ and $v_i$ as the ADM approximations
to $E(B_i|y)=E(\frac{V_i}{V_i+A}|y)$ and to $\operatorname
{Var}(B_i|y).$ Assume $E(\hat\beta_A|y)\doteq\hat\beta\equiv
\hat\beta_A$ and $E(P_A|y)\doteq P_{\hat A}$.
Then for $i=1,\ldots,k$
%
%
\begin{eqnarray}
E(\theta_i|y)&\doteq& (1-\hat B_i)y_i+\hat
B_ix_i^{\prime}\hat\beta\equiv\hat\theta_i,\label{eqgenmodtheoremexp}\\
\label{eqgenmodtheoremvar}\operatorname{Var}(\theta_i|y)&\doteq& \bigl(1-(1-p_{i,i})\hat B_i
\bigr)V_i
\nonumber
\\[-8pt]
\\[-8pt]
\nonumber
&&{}+v_i(y_i-\hat y_i)^2.
\end{eqnarray}
Here $p_{i,i}$ is the $i$th diagonal term in $P_{\hat A}$.
\end{theorem}

\begin{pf} Equation \eqref{eqgenmodtheoremexp} follows from
\eqref{eqgenmoddistrthetaintegrateoutbeta}, \eqref{eqgenmodexpthetaiAy} and $E(\hat\beta_A|y)=\hat\beta$,
since
\[
E(\theta_i|y)=E\{(1-B_i)y_i+B_i\hat y_i\}|y.
\]
%
Now use EVE's law (total variation) to get,
from~\eqref{eqgenmoddistrthetaintegrateoutbeta} and \eqref{eqgenmodvarexpthetaiAy},
%
%
\begin{eqnarray}
\qquad\operatorname{Var}(\theta_i|y)&=&E\operatorname{Var}(\theta
_i|A,y)+\operatorname{Var}(E\theta_i|A,y)\\
&=&E\{(1-B_i)V_i+B_ip_{i,i}V_i\}|y
\nonumber
\\[-8pt]
\\[-8pt]
\nonumber
&&{}+v_i(y_i-\hat
{y}_i)^2,
\end{eqnarray}
which is \eqref{eqgenmodtheoremvar}.

In our experience, these regression approximations when
$\pi(A)=A^{c-1},$ and $c=1$ especially, have been quite
satisfactory.
\citet{tang02}
provides a basis for making more precise approximations to
$E(P_A|y)$\vspace*{2pt} and to $E(\hat\beta_A|y)$ based on matrix and
determinant derivatives.
In the equal variance case, the
theorem's two moments are exact provided exact formulas for\break $E(B_i|y)$
and $\operatorname{Var}(B_i|y)$ are used.
However, Normality of
$\theta_i|A,y$ does not hold exactly for $\theta_i$ after
averaging over $A|y$, although that Normal approximation is commonly
made. 
\end{pf}

\section{Approximation Accuracy}\label{secapproxaccuracy}
\subsection{Approximation Accuracy of Shrinkages and the Random
Effects}\label{subsecapproxaccuracybetatheta}
Figures~\ref{figBvsT} and \ref{figBvsvhat}
show in the equal variance setting that
even for small samples like $k = 4, 10, 20$, the ADM approximation of
the first two exactly computed posterior moments of $B$ is quite good.
Our end goal, however, is verifying this
leads to good approximations of the posterior means
and variances of each random effect ($\theta_i, i = 1, \ldots, k$).

First, in the equal variance situation with $r = 0$,
we compare the weighted average of
posterior mean squared error of the $\theta_i$ values
via the ADM approximation
with this measure with the ``exact'' posterior mean.
Let us measure the difference of their mean squared errors,
given the data $y$, by computing
%
\begin{equation}
E\Biggl\{ \sum_{i=1}^k (\hat{\theta}_i - \theta_i)^2 | y\Biggr\}
\end{equation}
for the ADM approximation, with the expectation calculated exactly,
when $\pi(A) = 1$. Now
\begin{eqnarray*}
& & E\Biggl\{ \sum_{i=1}^k (\hat{\theta}_i - \theta_i)^2|y\Biggr\} \\
&&\quad =  E\Biggl\{\sum_{i=1}^k (\hat{\theta}_i - \hat{\theta}_{e,i})^2
\vert
y\Biggr\}
+ E\Biggl\{\sum_{i=1}^k (\hat{\theta}_{e,i} - \theta_i)^2 \vert y\Biggr\}\\
&&\quad =  \sum_{i=1}^k (\hat{\theta}_i - \hat{\theta}_{e,i})^2 +
\sum_{i=1}^k s_{e,i}^2,
\end{eqnarray*}
where the subscript $e$ denotes estimates done exactly (see Section~\ref{subsecexactmomentseqvar}),
with $s_{e,i}^2$ is given in \eqref{eqvarthetahatexact}.
Therefore
%
%
\begin{equation}
\label{eqpropvartheta}
\frac{\sum_{i=1}^k (\hat{\theta}_i -
\hat{\theta}_{e,i})^2}{\sum_{i=1}^k s_{e,i}^2}
\end{equation}
measures how well the ADM approximation works for random effects
estimates, smaller values indicating better approximations.
The highest (worst) ratio is $1.1\%$ which occurs
for $k$ near 20, and for 60\% shrinkage.
Greater accuracy holds for $k < 20$ and for $k > 20$. Thus,
in the equal variances setting, the
conditional mean squared errors of the ADM approximation
and the exact estimator of $\bold\theta$
never differ by more than $1.1\%$.

\begin{figure*}

\includegraphics{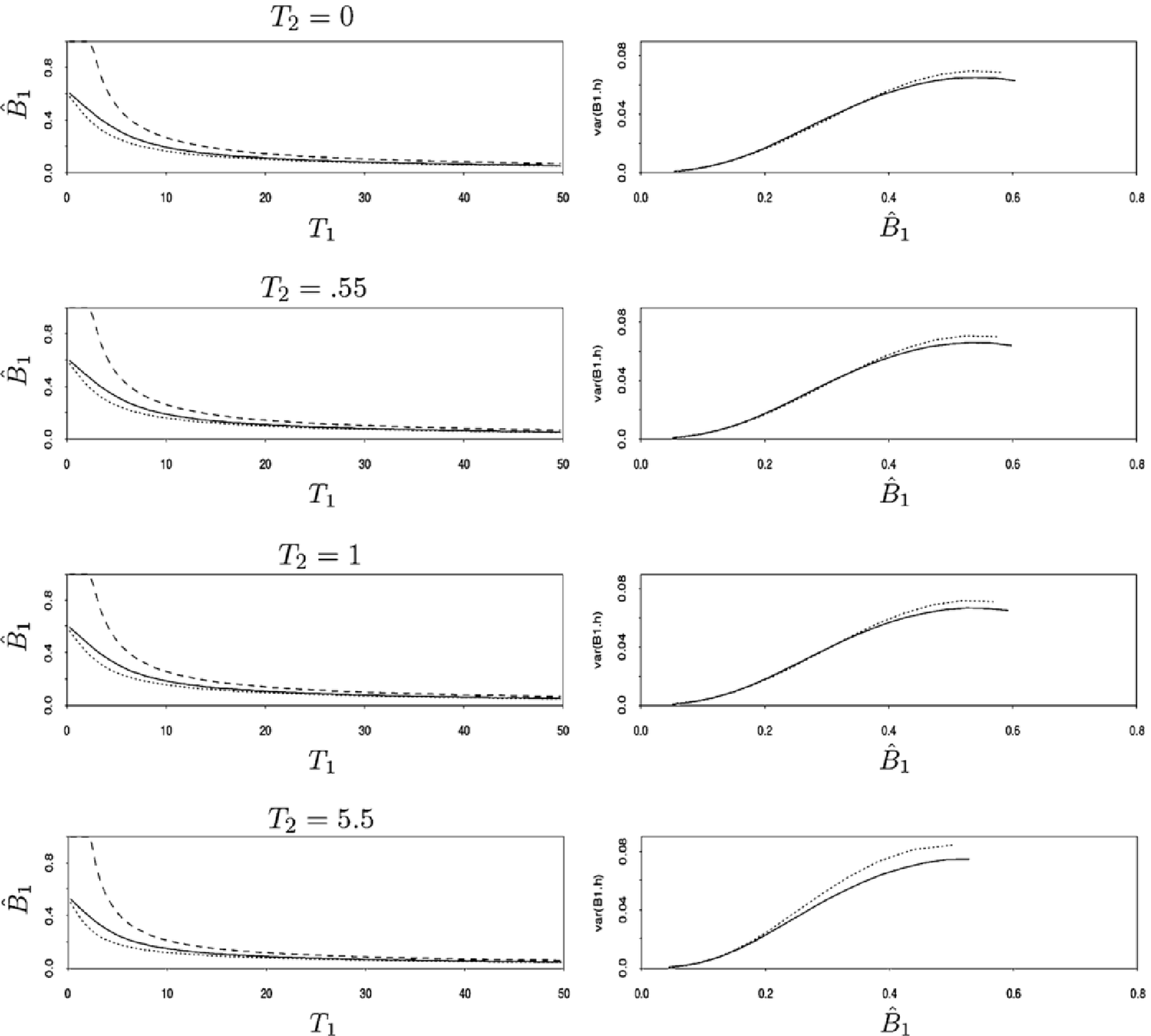}%
\vspace*{-3pt}
\caption{Approximation
accuracy for two groups of
variances, here for the small variance group\vspace*{1.5pt} ($k=10$, $r=1$,
$V_1=\cdots=V_5=0.55$, $V_6=\cdots=V_{10}=5.5$).
The left-hand side plots $\hat{B}_1$ against
$T_1$, with $T_2$ fixed at various values (which correspond
to $A=0, 0.55, 1, 5.5$). The right-hand side plots
$\var(B_1|\data)$ against $\hat{B}_1$.
Solid line is from the exact
method, dotted line from ADM approximation, long dashed
line is MLE.}
\label{figB1vsTunequal}
\vspace*{-7pt}
\end{figure*}


%
Now, still with $\pi(A)=1$, consider the unequal variance case
and ADM's accuracy for approximating the exact Bayes estimator
of $\theta$. The following example involves
two groups of variances for the~$y_i$ values, and
estimates the unknown mean vector $\mu_1 = \cdots= \mu_{10}$
in the second level (so $k=10, r=1$).
Five ``small'' variances are
set at \mbox{$V_1\!=\!V_2\!=\! \cdots\!=\! V_5 \!=\! 0.55$}, and five ``large'' ones
at $V_6\!=\!\cdots\!=\!V_{10}\!=\!5.5$.
Their~ma\-ximum-to-minimum variance ratio is a factor of~10, and their
harmonic mean is 1.0 (for convenience only).
Shrinkages $B_1 = \cdots= B_5 < B_6 = \cdots= B_{10}$ are toward
the nine-dimensional subspace orthogonal to the unit vector.
We calculated exact and
ADM means and variances of these shrinkages, which depend
on the separate values of the two-dimensional
statistic $T_1, T_2$ (these two sums of squares are standardized
by their respective~$2V_i$, each summed over its respective subgroup of
size~5, both centered on their common fitted grand mean).

\begin{figure*}

\includegraphics{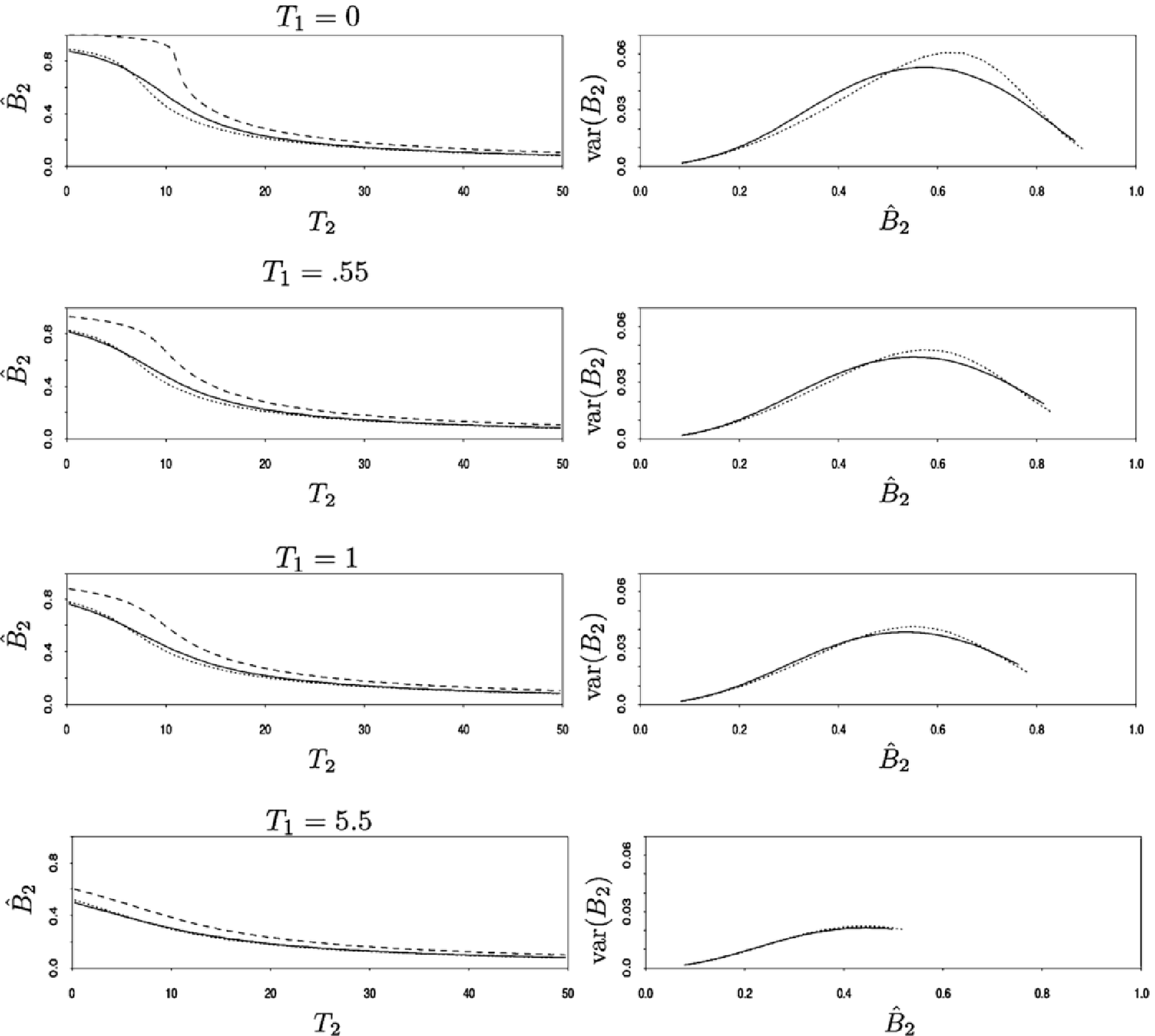}%
\vspace*{-3pt}
\caption{Approximation accuracy
for two groups of
variances, here for the large variance group ($k=10$, $r=1$,
$V_1=\cdots=V_5=0.55$, $V_6=\cdots=V_{10}=5.5$).
The left-hand side plots $\hat{B}_2$ against
$T_2$, with $T_1$ fixed at various values (which correspond
to $A=0, 0.55, 1, 5.5$). The right-hand side plots
$\var(B_2|\data)$ against $T_2$.
Solid line is from the exact
method, dotted line from ADM approximation, long dashed
line is MLE.}
\label{figB2vsTunequal}
\vspace*{-6pt}
\end{figure*}

Figure~\ref{figB1vsTunequal} concerns shrinkages for the first
five components with small variances,
$V_j = 0.55$, and Figure~\ref{figB2vsTunequal}
shows shrinkages for the five components with
large variances, $V_j = 5.5$.
The left panels of each figure show
shrinkage factor patterns for three different rules:
the MLE (dashed curve), the exactly
computed shrinkage using the harmonic prior for
which~$A$ has a flat density (solid curve),
and the ADM approximations to that
shrinkage factor (dotted curve). These are
graphed as a function of $T_1$ (Figure~\ref{figB1vsTunequal})
and $T_2$ (Figure~\ref{figB2vsTunequal}) with separate
displays, each conditional on one of four different
values of the opposite $T_j$.

Both figures show that the MLE has quite large shrinkages, just as
for equal variances.
The relationship between the ADM approximation and
the exactly computed expected
shrinkage that the ADM approximates is
similar to what was seen in the equal variance case.
The right-hand panels of each figure show
good agreement between the ADM
variance approximation and the exactly computed
variances $v_i$ when each is plotted against its own shrinkage.
The maximum shrinkages for ADM and the exact rule
are limited to values $<1$, curtailing the
horizontal axes for plots of $\operatorname{Var}(B_i)$.

To summarize for the prior $\pi(A)=1$,
the ADM approximations of exact shrinkage factors
for posterior means and variances of shrinkage factors
are slightly conservative, but generally are in good agreement with the
exact values obtained in the equal variance case. Similar results
hold for the unequal variance case when
variances $V_i$ differ by a factor of 10 and when $r=1$.

\vspace*{-2pt}

\section{Coverage Probabilities and Risk Functions}\label{seccoverageprobrisk}
\vspace*{-2pt}

Confidence interval coverage rates for $\theta_i$ are evaluated
next for the two main procedures of Section~2, both based on
assuming $A>0$ has a flat prior $\pi(A) = 1$ so that
the posterior density is the likelihood function.
One procedure, labeled ``exact'' here,
evaluates the exactly computed posterior means and
variances of $\theta_i$, given $y$, as in (34) and (35) for the equal
variances case, and otherwise by numerical integration.
It then assigns a Normal distribution with these two moments to
determine a posterior interval.
The second approach uses Normal distributions in the same way, but
centered and scaled via the ADM approximations of these two moments
in (44) and (45), or when $r \ge0$ and with unequal variances,
as in (57) and (58).
Normal distributions are not exact for~%
$\theta_i$, since the actual distributions
are skewed (right-skewed for relatively large $y_i$, and
left-skewed for small $y_i$).
This matters less in repeated sampling
evaluations that randomize over $y$, making skewnesses average
to zero for each~$i$.

\begin{figure*}

\includegraphics{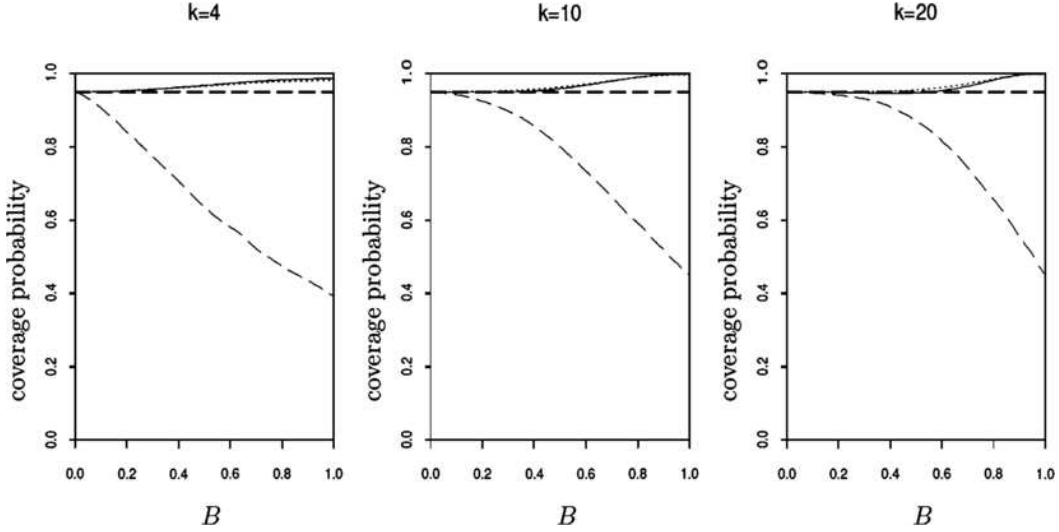}

\caption{Plot of coverage probabilities of $\theta_i$ (random
effects) for
equal variances without regression, each against the true
shrinkage factor. In the three graphs ($k=4,10,20$), both
the ``exact'' Bayes (solid curve) and its ADM-SHP approximation
(dotted curve) achieve approximately the nominal
$0.95$ coverage rates (as indicated by the bold dashed horizontal
line), or higher. The MLE (long dashes)
can be markedly nonconservative,
especially with large true shrinkages $B$ ($A$ near $0$).
As $A$ approaches $0$,
MLE coverages fall below $50\%$, however large $k$ might be.}
\label{figcoveragetheta}
\end{figure*}

For all $i = 1, \ldots, k$, we seek two-tailed
frequency coverage probabilities as a function of $A$:
%
%
\begin{eqnarray}
\Pr\biggl[ \frac{(\theta_i - \hat{\theta}_i)^2}{s_i^2} \le
(z^\ast)^2 | A \biggr],
\end{eqnarray}
when the nominal coverage is $95\%$, so $z^\ast= 1.96$.
Each procedure studied uses its own estimate $s_i^2$ of
the conditional variance of $\theta_i$.
A related measure directly assesses how well each $s_i^2$
envelops the expected squared error, given $A$, with values $ \le1$
indicating that $s_i$ assigns sufficiently large intervals:
%
%
\begin{equation}
\label{eqrisk2}
E \{ (\theta_i-\hat{\theta}_i)^2 / s_i^2 | A \}.
\end{equation}
%

Details of the simulation are in \citet{tang02}, where
Rao--Blackwellization increased
the accuracy
by evaluating some conditional Normal distributions exactly, given
$A$ and $y$.
That is, for \eqref{eqrisk2},
\begin{eqnarray*}
& & \Pr\biggl\{\frac{(\hat{\theta}_i - \theta_i)^2}{s_i^2} \le
(z^\ast)^2 | A \biggr\} \\
&&\quad =  E \biggl[\Pr\biggl\{\frac{(\hat{\theta}_i -
\theta_i)^2}{s_i^2} \le(z^\ast)^2 | y, \beta, A
\biggr\} \Big| A \biggr] \\
&&\quad =  E \biggl[\Phi\biggl\{\frac{\hat{\theta}_i - (1-B_i) y_i -B_i x'_i
\beta+ z^\ast s_i}{\sqrt{V_i (1-B_i)}}\biggr\}\\
&&\qquad{}- \Phi\biggl\{\frac{\hat{\theta}_i - (1-B_i) y_i - B_i x'_i \beta
-z^\ast s_i}{\sqrt{V_i (1-B_i)}}\biggr\}\Big| A \biggr].
\end{eqnarray*}


\begin{figure*}

\includegraphics{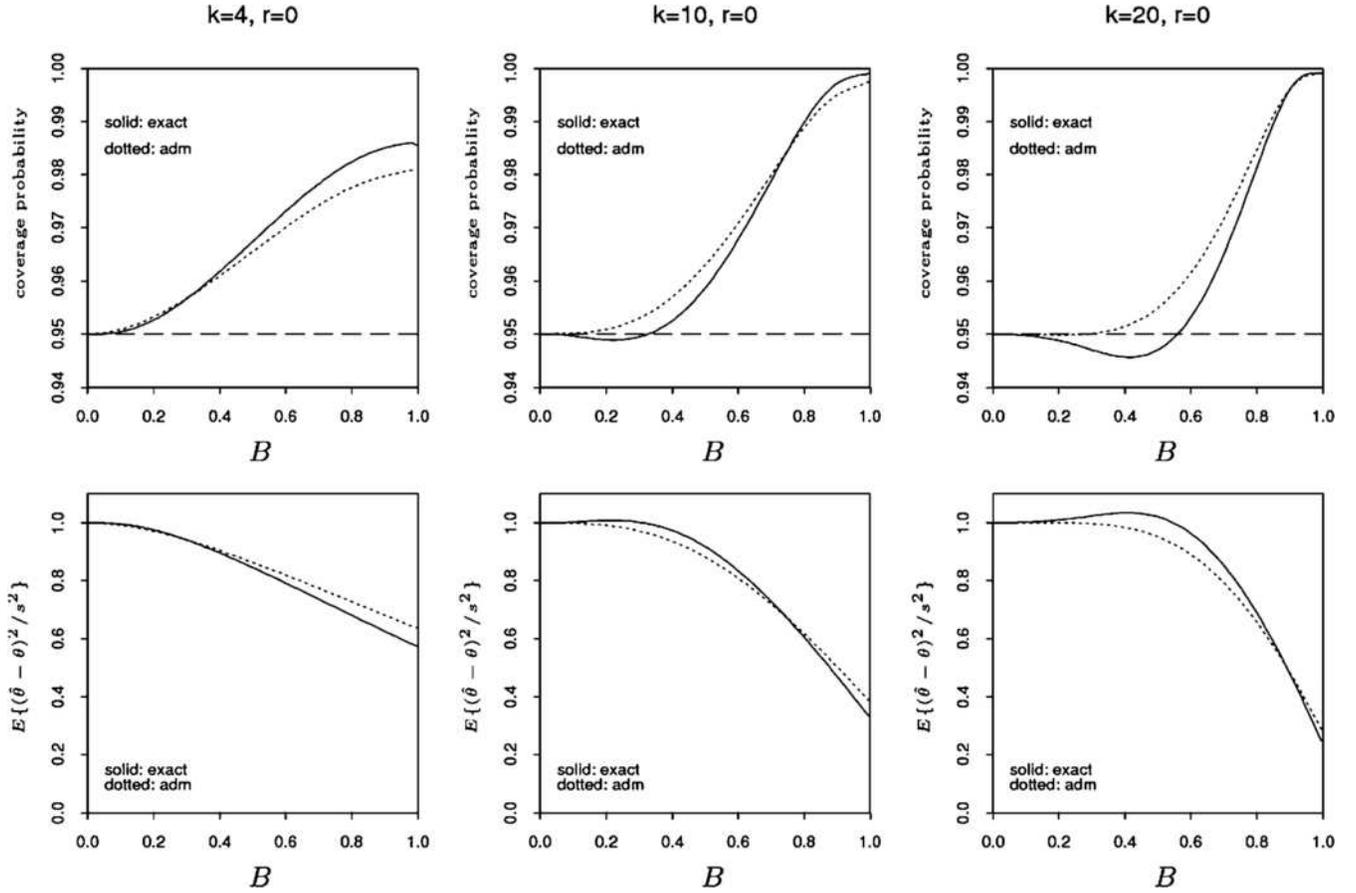}

\caption{Plot of coverage
probabilities and standardized risk functions for equal variances
without regression. The first row plots coverage
probabilities against true values of $B = V/(V+A)$
on a larger scale than in Figure~\protect\ref{figcoveragetheta},
with ADM-SHP coverages being the dotted curves. The
second row plots the expected value of the loss function
calibrated by $s_i^2$ \protect\eqref{eqrisk2}.}
\label{figcoveragestdrisktheta}
\end{figure*}

\subsection{Equal Variances Example}

Figure~\ref{figcoveragetheta} plots the actual coverage probabilities
for the three confidence interval procedures, each against the possible
``true'' $B$ values, for three equal variance procedures always
with $r = 0$, and for
$k=4$ ($m=1$), $k=10$ ($m=4$) and $k=20$ ($m=9$). For each
$B=0.005, 0.015, \ldots, 0.995$, $1000$ data sets were generated and
the interval procedures for ``exact,'' its ADM approximation, and the
MLE were
evaluated and averaged to estimate the coverage probabilities.
Confidence intervals for the MLE were determined
simply by taking each variance to be the MLE $V (1 - B)$.
These MLE coverages are plotted with long dashes
in Figure~\ref{figcoveragetheta}.
When shrinkage $B$ is large, these MLE intervals give poor coverages,
ultimately dropping to just under $50\%$, as shown in Section 2.

The graph of Figure~\ref{figcoveragetheta} is redone in the first row
of Figure~\ref{figcoveragestdrisktheta},
but without the MLE. That allows an amplified scale that shows the
slight differences in coverage rates between
the ``exact'' rule and its ADM-SHP approximation.
The ADM-SHP coverages meet or exceed
$\ge0.95$ for all $A$ (within simulation error).
The ``exact'' procedure's coverages can be slightly nonconservative,
but its lowest coverage is at least $0.945$ (when $k=20$ and $B=0.4$)
for all $k$ shown. The ADM-SHP intervals achieve (or exceed) their
nominal $0.95$ coverage rates by having slightly wider intervals
than ``exact,'' due to ADM's reduced
shrinkage estimates and its larger
variance estimates $v$, as studied in Section~3.
As $B$ increases both methods become
quite conservative, with coverages well above $0.95$.

The bottom row of Figure~\ref{figcoveragestdrisktheta} plots the
function~\eqref{eqrisk2} against $B$ to
compare the two different methods. Values less than 1.0
indicate that the estimated variances $s_i^2$
average to as much as or more than the average mean square.
This further suggests that the interval coverages will (nearly)
provide the nominal coverage ($95\%$) for all values of $A>0$.

\subsection{An Unequal Variances Example: Two Groups of Variances}\label{subsecunequalvarexample}

We return to the unequal variances example of Section 3 with $k=10$,
$r=1$, $V_1=\cdots=V_5=0.55$, and $V_6=\cdots=V_{10}=5.5$.
For this simulation, $100$ data sets were generated for each
of 50 values $B_0 =0.01, 0.03, \ldots, 0.99$, where $B_0 \equiv V_0/(V_0+A)$
and $V_0=1$ is the harmonic mean of the $V_i$.
Nominal $ 95\%$ confidence intervals for each $\theta_i$ were
evaluated for each data set.
The confidence rates and average calibrated losses \eqref{eqrisk2} then
were averaged over the simulated values.

\begin{figure}

\includegraphics{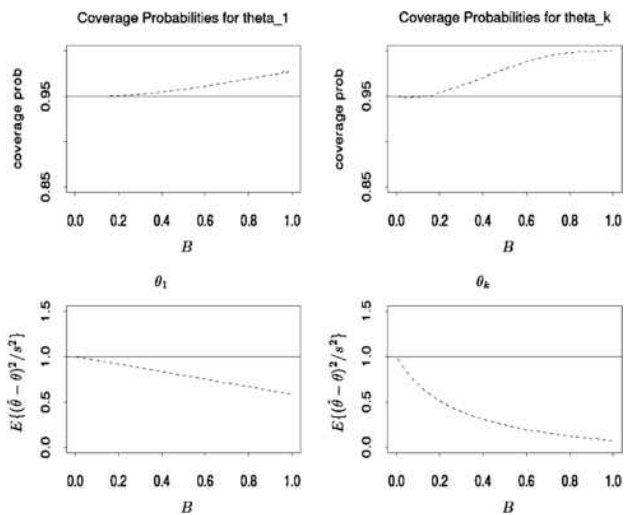}

\caption{Plot of
ADM-SHP coverages and expected value of average
calibrated losses against $B_0 = 1/(1+A)$. Here
$k=10, r = 1$, $V_1=\cdots=V_5=0.55$, $V_6 = \cdots = V_{10} = 5.5$.}
\label{figcoveragethetaunequal}
\end{figure}

Figure~\ref{figcoveragethetaunequal} plots coverages
of the ADM-SHP intervals
and calibrated risk functions \eqref{eqrisk2}
for $\theta_1$ and for $\theta_{10}$ as $B_0 = 1/(1+A)$ varies.
The upper left panel of Figure~\ref{figcoveragethetaunequal} plots the
coverage probabilities against $B_0$ for the group of five with
small variances $V_i = 0.55$,\vadjust{\goodbreak} and the upper right for the
remaining group of five with large variances $V_i = 5.50$.
As $B_0$ increases and $A$ decreases, coverage rates
generally increase. Coverages achieve or exceed
their nominal $0.95$ levels (within simulation error),
while for small $A$ and big~$B_0$, coverages for the
large variance group substantially exceed both their nominal rate and
the coverages for the small variance group.
The calibrated risks are less than 1.0
in Figure~\ref{figcoveragethetaunequal}
which show that the intervals are wide enough to be conservative,
although they may be excessively conservative for the large variance group.
One remedy could be using the scale-invariant prior $c=0.5$, which
makes $\sqrt A$ flat.
Coverages rates for the exact version of SHP were not evaluated
for this unequal variance case, and that can be time-consuming for
repeated sampling.
Simple and fast computing, plus a procedure's transparency, are
reasons for finding simple and accurate approximations.

\section{Conclusions}\label{secconclusions}

Why might a Bayesian or objective Bayesian sta\-tistician
who has settled on prior distribution $\pi(A)$ on $A$
consider approximating with ADM?
There are several reasons, beyond the general observation that
any procedure used in an application is an approximation.

\begin{enumerate}[1.]


\item[1.]
Speed of convergence is valuable with big data sets,
especially if a procedure is to be used repeatedly
for model selection and model checking.
The approximations here avoid MCMC burn-ins.
Speed also makes
it feasible to simulate many times, for example, for bootstrapping, or to
check a procedure's operating characteristics.

\item[2.] Data analysts may need to obtain the same
results each time a particular model is re-fit to the same data,
which stochastic approximations do not do.


\item[3.] MLE methods always will play a central role in
statistics. For the model of this paper, ADM maintains the spirit
of MLE while making small sample improvements.

\item[4.] Using ADM to help fit shrinkage factors extends to
multilevel generalized linear models, for example, to fit
a Poisson model (\citep{chrimorr97}). In such more complicated
non-Normal models, MCMC and exact numerical integration may be
more difficult or impossible, giving MLE and ADM a greater
advantage of ease. Then the frequency properties of ADM
can be checked with each data application by simulating or
bootstrapping from the fitted multilevel model. However, that
will not reveal how well ADM approximates the exact Bayes procedure.

\item[5.] Multiplying a likelihood by $A$ before maximizing
combines neatly with EM methods as
used to find the MLE of $A$ (\citep{demplairrubi77}).
With ADM, EM would avoid infinite loops that occur
when the MLE $\hat{A} = 0$.

\item[6.] Data analysts always will need well-checked, pre\-packaged, documented,
widely known and available procedures for fitting models.

\item[7.] Statistical software programmers should find it easy to
program and adopt the ADM-SHP formulas, for example, the formulas
of Section 2.8, in standard software.
For example, ADM could be an option in SAS PROC MIXED
along with MLE and REML.
\end{enumerate}

Barring prior information that $A$ is likely to be small,
the ADM-SHP methods developed here for~ma\-king inferences,
especially interval estimates, about
the random effects in a two-level Normal regression model
will have better frequency performance
over the entire range of $A \ge0$ than MLE and REML methods.
Our derivation has benefited from viewing
Stein's harmonic prior SHP on the random effects~$\theta_i$
as arising from a uniform mixture over $A$ of the Level-2
Normal distribution $\eqref{eqlevel2}$, 
that is, according to $\pi(A) = 1$.

With this formal (improper) prior, the posterior density on $A$
agrees with the marginalized likelihood function $L(A)$.
That justifies the term ``adjustment for likelihood maximization''
when ``ALM'' is restricted to point estimation of a shrinkage factor.
The results here go on to use
the flat $\pi(A) = 1$ prior and conditional (Bayesian) reasoning
as a guide to accounting for variability
of the shrinkage factors~$B_i$
and ultimately, of the random effects $\theta_i$.
ADM approximates the exact Bayes procedures with
considerable accuracy, given that it retains the (relative) ease of
MLE/REML calculations, that is, by using two derivatives
of the adjusted log-likelihood\break log($A$ $L(A)$).
Of course the adjustment here more generally
would adjust by using the multiplier $\pi(A)$ if $\pi(A) \ne1$.
While more testing is needed for unequal variances cases,
the confidence intervals for random effects
arising from the ADM-SHP combination here thus far have met or
exceeded their nominal coverages if $k-r \ge3$.
Still, the search should continue
for priors on $A$ that will provide even better frequency
interval coverages.

\section*{Acknowledgments}
The authors
gratefully acknowledge funding for this project provided in part by
NSF Grant DMS-97-05156,\ and for many helpful
suggestions made by the Editors, the Associate Editor and
a referee.

%

\end{document}